\documentclass{article}
\usepackage{graphicx} 
\usepackage{amsfonts,amsmath,amssymb}
\numberwithin{equation}{section}
\usepackage{tabularx}
\usepackage{booktabs}
\usepackage{todonotes}
\usepackage{geometry}
\usepackage{tcolorbox}
\usepackage{caption,subcaption}
\usepackage[ruled,vlined]{algorithm2e}
\usepackage{amsmath}
\usepackage{multirow}
\usepackage{caption}
\usepackage{subcaption}
\usepackage{hyperref}
\usepackage{tabularx}
\usepackage{mathtools}
\usepackage[utf8]{inputenc}
\usepackage{authblk}

\DeclarePairedDelimiterX\braket[2]{\langle}{\rangle}{#1 \delimsize\vert #2}
\DeclarePairedDelimiterX\ketz[1]{\vert}{\rangle}{\mathcal{Z}^{\,4\,#1}}
\DeclarePairedDelimiterX\braz[1]{\langle}{\vert}{\mathcal{Z}^{\,#1\,4}}
\DeclarePairedDelimiterX\comm[2]{[}{]}{#1 \,,\delimsize #2}
\DeclarePairedDelimiterX\anticomm[2]{\lbrace}{\rbrace}{#1 \,,\delimsize #2}
\DeclarePairedDelimiterX\kete[1]{\vert}{\rangle}{e^{#1}}
\DeclarePairedDelimiterX\brae[1]{\langle}{\vert}{e^{#1}}
\newcommand{\zv}[1]{\mathcal{Z}^{\,4\, #1}}
\newcommand{\zb}[1]{\mathcal{Z}^{\,#1\, 4}}

\newcommand{\bI}{\mathbb{I}}
\DeclarePairedDelimiterX\inner[2]{\langle}{\rangle}{
\zb{#1} \delimsize\vert \zv{#2}}

\DeclarePairedDelimiterX\kzh[1]{\vert}{\rangle}{\mathbf{z}^{#1}}
\DeclarePairedDelimiterX\bzh[1]{\langle}{\vert}{\mathbf{\tilde{z}}^{#1}}
\DeclarePairedDelimiterX\innerh[2]{\langle}{\rangle}{
\mathbf{\tilde{z}}^{#1} \delimsize\vert \mathbf{z}^{#2}}

\newcommand{\tr}{\mathop{\rm Tr}}

\newcommand{\cZ}{{\cal Z}}

\DeclarePairedDelimiterX\ket[1]{\vert}{\rangle}{#1}
\DeclarePairedDelimiterX\bra[1]{\langle}{\vert}{#1}

\usepackage{tikz}
\usepackage{xparse}

\tikzset{plaqLine/.style={blue, line width=0.1pt}}
\tikzset{plaqPathbr/.style={brown, line width=3pt, opacity=0.3}}
\tikzset{plaqPathbl/.style={blue, line width=3pt, opacity=0.3}}

\NewDocumentCommand{\plaquetteEq}{O{1.0} m m m m m}{%
  \mathord{%
    \begin{tikzpicture}[baseline={(0,0)}, scale=#1]%
      \draw[plaqLine] (0,0) -- (0.5,0.5);
      \draw[plaqLine] (0.5,0.5) -- (1,0);
      \draw[plaqLine] (1,0) -- (0.5,-0.5);
      \draw[plaqLine] (0.5,-0.5) -- (0,0);

      \IfValueT{#2}{%
        \fill[#2,opacity=0.1]
          (0,0) -- (0.5,0.5) -- (1,0) -- (0.5,-0.5) -- cycle;
      }

      \IfValueT{#3}{\node[anchor=east,  font=\scriptsize]  at (0,0)      {$#3$};}
      \IfValueT{#4}{\node[anchor=south, font=\scriptsize]  at (0.5,0.5)  {$#4$};}
      \IfValueT{#5}{\node[anchor=west,  font=\scriptsize]  at (1,0)      {$#5$};}
      \IfValueT{#6}{\node[anchor=north, font=\scriptsize]  at (0.5,-0.5) {$#6$};}
    \end{tikzpicture}%
  }%
}

\NewDocumentCommand{\dplaquetteEq}{O{0.5} m m m m m}{%
  \mathord{%
    \begin{tikzpicture}[baseline={(0,0)}, scale=#1]

      \draw[plaqLine] (0,0) -- (0.5,0.5);
      \draw[plaqLine] (0.5,0.5) -- (1,0);
      \draw[plaqLine] (1,0) -- (0.5,-0.5);
      \draw[plaqLine] (0.5,-0.5) -- (0,0);

      \IfValueT{#2}{%
        \fill[#2,opacity=0.1]
          (0,0) -- (0.5,0.5) -- (1,0) -- (0.5,-0.5) -- cycle;
      }

      \IfValueT{#3}{\node[anchor=east,  font=\scriptsize]  at (0,0)      {$#3$};}
      \IfValueT{#4}{\node[anchor=south, font=\scriptsize]  at (0.5,0.5)  {$#4$};}
      \IfValueT{#5}{\node[anchor=west,  font=\scriptsize]  at (1,0)      {$#5$};}
      \IfValueT{#6}{\node[anchor=north, font=\scriptsize]  at (0.5,-0.5) {$#6$};}
      
      \begin{scope}[shift={(0.5,0.5)}]
      \fill (-0.07,0) arc[start angle=180, end angle=360, radius=0.07] -- cycle;
      \end{scope}

      \begin{scope}[shift={(0.5,-0.5)}]
      \fill (0.07,0) arc[start angle=0, end angle=180, radius=0.07] -- cycle;
      \end{scope}
    \end{tikzpicture}%
  }%
}

\def\cZ{\mathcal{Z}}

\newcommand{\Tr}{{\rm Tr}}
\def\tr{\mathop{\mathrm{tr}}}

\title{\boldmath Machine Learning Topological Order from Defect Partition Functions}

\author[1]{Kazem Bitaghsir Fadafan}
\author[2]{Wei Cui}
\author[2,3]{Babak Haghighat}
\author[2]{Shailesh Lal}
\affil[1]{Faculty of Physics, Shahrood University of Technology, 
P.O.Box 3619995161 Shahrood, Iran}
\affil[2]{Beijing Institute of Mathematical Sciences and Applications (BIMSA) Huaibei Town, Huairou District, Beijing 101408, China}
\affil[3]{Yau Mathematical Sciences Center, Tsinghua University, Beijing, 100084, China}
\date{}
\begin{document}

\maketitle
\begin{abstract}
We introduce a machine learning framework for extracting Ising topological order from defect partition functions of the two-dimensional Ising model on a torus. Restricted Boltzmann Machines (RBMs) are trained on Ising model sampled at criticality across topological sectors. We take a component-wise square-root map of the learned distributions which naturally produces candidate wavefunctions for the (2+1)-dimensional Ising TQFT. As a nontrivial consistency check, we extract the modular S-matrix from overlaps of the resulting states and recover the expected Ising modular data. Our results demonstrate that neural network representations can capture both critical fluctuations and emergent topological structure, providing a data-driven route from lattice statistical mechanics to topological quantum field theory.
\end{abstract}
\tableofcontents
\section{Introduction \& Overview}
String-net condensation has been proposed as a fundamental physical mechanism for realizing topological phases, as discussed in \cite{Levin:2004mi}. This framework provides a unifying perspective on quantum many-body systems with long-range entanglement, playing a crucial role in understanding topologically ordered states, such as those appearing in fractional quantum Hall systems and certain quantum spin liquids.

One well-known example of a dual description in this context is the duality between $Z_2$ lattice gauge theory and the Ising model in (2+1) dimensions \cite{kogut1979}. This duality reveals deep connections between gauge theories and statistical mechanics, where the ordered phase of the Ising model corresponds to the deconfined phase of the $Z_2$ gauge theory, while the disordered phase of the Ising model corresponds to the confined phase in the gauge theory. Additionally, string-like excitations in the gauge theory naturally map to domain walls in the Ising model.

This relationship provides insight into confinement mechanisms, quantum phase transitions, and emergent degrees of freedom in lattice models, making it an essential tool in condensed matter physics and quantum information science. Furthermore, these ideas extend to modern applications in quantum error correction, tensor network representations, and deep learning approaches to gauge theories.

We are interested in the profound TQFT/CFT correspondence, which establishes a deep connection between Topological Quantum Field Theories (TQFTs) in (2+1) dimensions, describing the low-energy physics of topological phases through features like topology-dependent ground state degeneracy and anyonic excitations, and Conformal Field Theories (CFTs) in (1+1) dimensions, which describe universal critical points and crucially provide the edge theories of topological phases. It was shown that the states in the Hilbert space of a Chern-Simons TQFT can be represented by conformal blocks of a related CFT.

Monte Carlo methods have significantly advanced lattice gauge theory, but challenges like the sign problem remain. Neural network quantum states offer an alternative, accurately capturing ground states and studying topological phases and confinement transitions in lattice $Z_N$ gauge theories \cite{Apte:2024vwn}.

Restricted Boltzmann Machines (RBMs) for finding groundstates of spin lattice models were first advocated in \cite{Carleo:2016svm}. Using an RBM, a lattice Fractional Quantum Hall state has been studied in \cite{Glasser:2018bzy}. The Moore-Read states on a lattice and parent Hamiltonian is obtained in \cite{Glasser:2015fla,Manna:2018xve}. An RBM architecture has also been applied to find groundstates of Kitaev's Honeycomb model which is yet another lattice manifestation of the Ising TQFT \cite{Noormandipour:2020dqp}. However, in all these cases one encounters the problem that the gradient flow to minimize energy expectation values is too slow when the lattice size grows. In most cases this is due to network parameters approaching a local minimum. Another subtlety is that in chiral topological phases, the RBM has to learn complex parameters which apart from the wavefunction amplitudes also include their phases. 

In this paper we want to advocate a novel approach by disentangling these two types of parameters and first learning the amplitudes. This way, we learn a certain instance in the universality class of the TQFT. Other instances can be obtained by subsequently training the network to learn wavefunction phases based on a given concrete lattice Hamiltonian. Our general philosophy is as follows. To learn the groundstates of a given 3d TQFT, we first identify a 2d lattice model which at criticality flows to the boundary CFT. We then use Monte Carlo sampling to obtain snapshots of spin configurations at the critical point and subsequently train an RBM based on these snapshots. A simple procedure then allows us to map these to groundstate wavefunctions. We apply this procedure to the 2d Ising model.

The paper is organized as follows. Section \ref{sec:2dIsing} reviews the 2d classical Ising model at criticality on the torus \(T^2\) along with the insertion of line defects. Monte-Carlo simulations for the untwisted Ising model as well as defect insertions at criticality are discussed in Section \ref{sec:mcsim}. Section \ref{sec:IsingTQFT} presents the variational ansatz for the TQFT wave functions motivated by the RBM representation of the partition function. 
Section \ref{sec:IsingRBM} implements an RBM implementation of the Ising model at criticality along with defect insertions. Our analysis demonstrates how the RBM `effective action' captures physics of the critical Ising model. Finally Section \ref{sec:RBMWave} constructs the RBM ansatz wave functions obtained through the RBMs as proposed in the paper and studies some properties. We then conclude with a summary and outlook. The appendix provides technical details about training the RBM.

\paragraph{Notation and Conventions}
The geometric, field-theoretic, and machine learning conventions utilized throughout this work are consolidated in Table~\ref{tab:notation_conventions} for reference.
\begin{table}
\centering
\small
\caption{Summary of Lattice, CFT, and Restricted Boltzmann Machine Conventions}
\label{tab:notation_conventions}
\begin{tabular}{llp{7.5cm}}
\toprule
\textbf{Domain} & \textbf{Symbol} & \textbf{Remarks} \\
\midrule
\textbf{Lattice} 
& $\Lambda$ & 2d square lattice, sites indexed by $I \equiv (x_I, y_I)$. \\
& $x_I, y_I$ & toroidal lattice coordinates valued in \(\mathbb{Z}_M, \mathbb{Z}_L\) \\
& $a, b$ & non-contractible cycles of the discretized torus. \\
& $\tau = 1$ & modular parameter fixed by setting $M = L$. \\
& $\sigma_I \in \pm 1$ & on-site classical Ising spin degrees of freedom. \\
& $E$, $e \equiv \langle IJ \rangle$ & Edge set of \(\Lambda\), contains all nearest-neighbor links. \\
& $K_e, K_{IJ}$ & dimensionless interaction strengths. \\
& $\tilde{\Lambda}, \tilde{I}, \tilde{e}, K_{\tilde{e}}^*$ & dual lattice and parameters \\
\midrule
\textbf{CFT/TQFT} 
& $\{\mathbb{I}, \psi, \sigma\}$ & conformal primaries (Identity, Fermion, Spin). \\
& $\mathcal{Z}^{\mathcal{L}_3}_{\mathcal{L}_a \mathcal{L}_b}$ & CFT partition function on \(T^2\). \\
& $\mathcal{L}_3 \in \{\mathbb{I}, \psi, \sigma\}$ & conformal primary insertion \\
& $\mathcal{L}_a, \mathcal{L}_b \in \mathbb{Z}_2$ & symmetry-twisted b.c. / defect lines along $a, b$ cycles. \\
\midrule
\textbf{RBM/ML} 
& $G = (\mathcal{V} \cup \mathcal{H}, \mathcal{E})$ & Bipartite graph with visible and hidden nodes $\mathcal{V}$\, $\mathcal{H}$. \\
& $\mathbf{v} \in \mathfrak{V}$ & Visible node vector in $\{0,1\}^{n_v}$ or $\{-1,+1\}^{n_v}$. \\
& $\mathbf{h} \in \mathfrak{H}$ & Hidden/latent state vector in $\{0,1\}^{n_h}$ or $\{-1,+1\}^{n_h}$. \\
& $\boldsymbol{\theta} \equiv \{\mathbf{W}, \mathbf{b}, \mathbf{c}\}$ & RBM parameters mediating correlations. \\
& $\mathbf{W} \in \mathbb{R}^{n_h \times n_v}$ & mediates \(\mathbf{v},\mathbf{h}\) and hence
\(\mathbf{v},\mathbf{v}\) correlations\\
& $E(\mathbf{v}, \mathbf{h}; \boldsymbol{\theta})$ & Network energy function: $- \mathbf{h}^T\mathbf{W}\mathbf{v} - \mathbf{b}^T\mathbf{h} - \mathbf{c}^T\mathbf{v}$. \\
& $P(\mathbf{v}, \mathbf{h}; \boldsymbol{\theta})$ & Boltzmann joint pdf: $\mathcal{Z}_{\text{RBM}}^{-1} \exp(-E(\mathbf{v}, \mathbf{h}))$. \\
& $P(\mathbf{v}; \boldsymbol{\theta})$ & target density function $\sum_{\mathbf{h}} P(\mathbf{v},\mathbf{h})$ over data. \\
\bottomrule
\end{tabular}
\end{table}
\section{Ising Model on the two-Torus}
\label{sec:2dIsing}
\begin{figure}
\centering
\includegraphics[width=0.7\linewidth]{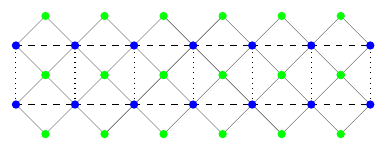}
\caption{Geometric configuration of the two-dimensional Ising model on the primary lattice $\Lambda$ (blue sites) and its corresponding dual lattice $\tilde{\Lambda}$ (green sites). Dashed and dotted lines denote horizontal and vertical edges on $\Lambda$, respectively. Each dual site $\tilde{p} \in \tilde{\Lambda}$ is positioned at the center of its associated primary plaquette, shifting it by $(+1/2, +1/2)$ relative to the primary site $p \in \Lambda$, which effectively rotates the dual edge axes by $45^\circ$ relative to the primary lattice.}
\label{fig:plaquettelattice}
\end{figure}

Consider the square lattice $\Lambda \subset \mathbb{Z}^2$ with sites indexed by coordinate pairs $I \equiv (x_I, y_I)$ and nearest neighbors connected by edges $e \equiv \langle IJ \rangle$. Plaquettes $\pi$ on $\Lambda$ are oriented closed loops 
\begin{equation}
\label{eq:plaquette_def}
\pi \quad : \quad (x, y) \rightarrow (x+1, y) \rightarrow (x+1, y+1) \rightarrow (x, y+1) \rightarrow (x, y) \,.
\end{equation}
The above plaquette is centered at the dual position $(x + \frac{1}{2}, y + \frac{1}{2})$. The canonical dual lattice $\tilde{\Lambda}$ is obtained by placing dual sites $\tilde{I}$ on these centers and linking them via dual edges $\tilde{e}_{IJ} \equiv \langle \tilde{I}\tilde{J} \rangle$ that are geometrically normal to the original edges $e \equiv \langle IJ \rangle$. We use \(E\) to denote the set of all edges \(e\) of \(\Lambda\). We identify the coordinate labels $x$ and $y$ modulo $M$ and $L$ respectively to obtain a toroidal lattice with $M$ rows and $L$ columns. Non-contractible cycles wrap along the vertical and horizontal directions of this torus. 
Figure~\ref{fig:plaquettelattice} 
illustrates this geometry, with blue sites on $\Lambda$ and green sites on 
$\tilde{\Lambda}$. 
Define an Ising model with spins $\sigma_I \in \{\pm 1\}$ living on the lattice sites $I$ and dynamics controlled by the Hamiltonian
\begin{equation}
\label{eq:ising_H}
H_{\mathrm{lattice}} = \sum_{\langle IJ \rangle} U_{IJ} \sigma_I \sigma_J \,,    
\end{equation}
where we impose periodic boundary conditions on the spins along both torus cycles.
Placing the system at temperature \(\beta^{-1}\) yields the partition function
\begin{equation}
\label{eq:ising_partition}
\mathcal{Z}_{\mathrm{lattice}} = \sum_{\{\sigma\}} \exp \left( \sum_{\langle IJ \rangle} K_{IJ} \sigma_I \sigma_J \right) \,,    
\end{equation}
where $K_{IJ} = -\beta U_{IJ}$ are dimensionless and encode both the physical Hamiltonian couplings $ U_{IJ}$ and the inverse temperature $\beta$. The Ising model \eqref{eq:ising_H} has two global symmetries which are especially relevant here. One is the \(\mathbb{Z}_2\) symmetry
obtained by flipping \(\sigma_I\to -\sigma_I\) at every site \(I\in\Lambda\) and the other is Kramers-Wannier duality \cite{PhysRev.60.252,PhysRev.60.263}\,. The duality transform maps the Ising model on the primary lattice \(\Lambda\) to an equivalent Ising model on \(\tilde{\Lambda}\) with dual couplings defined via:
\begin{equation}\label{eq:kw_longform}
\sinh 2\tilde{K}\left[\tilde{e}\right] = \frac{1}{\sinh 2K\left[e\right]}\,,
\end{equation}
where \(\tilde{e}\) is the dual lattice edge which intersects the primary lattice edge \(e\).
The dictionary \eqref{eq:kw_longform} follows from a careful analysis of the Ising partition function \eqref{eq:ising_partition}. This analysis reveals several highly nontrivial elements that make the Kramers-Wannier (KW) duality possible. First, the pairing of edges $e$ and $\tilde{e}$ required for the duality originates in kinematics. It follows from the observation that the contribution of a given edge $e = \langle IJ \rangle$ to \eqref{eq:ising_partition} is captured by the $\pm 1$-valued bond variables $\tau_{IJ} = \sigma_I \sigma_J$, which are subject to the non-local constraint
\begin{equation}
\prod_{e \in \partial \pi} \tau_e = 1\,,
\end{equation}
as may be readily seen by traversing the path \eqref{eq:plaquette_def}. This constraint can be trivialized by substituting
\begin{equation}
\tau_e = \mu_{\tilde{I}} \mu_{\tilde{J}}\,,
\end{equation}
such that $\tilde{e}=\langle\tilde{I}\tilde{J}\rangle$ is the dual lattice edge intersecting the original link $e$. This substitution explicitly recovers the correspondence between edges established in \eqref{eq:kw_longform}. The topology of the underlying lattice plays a non-obvious but crucial role, which becomes apparent when expanding \eqref{eq:ising_partition} as a sum over edges $e\in E$. Schematically, this yields
\begin{equation}
\mathcal{Z} 
= \sum_{E'\subseteq E}
\prod_{e\in E'}\tanh\left(K_e\right)
\left(\sum_{\{\sigma\}}
\prod_{\langle IJ\rangle \in E'}
\sigma_I\sigma_J\right)\,,
\end{equation}
where we have omitted all but the most essential factors. Let $d_I(E')$ denote the degree of vertex $I$ within the subset $E'$. The spin sum then factors as
\begin{equation}
\sum_{\{\sigma\}}
\prod_{\langle IJ\rangle \in E'} \sigma_I\sigma_J
=
\prod_I
\sum_{\sigma_I=\pm1}
\sigma_I^{d_I(E')}\,,    
\end{equation}
where $I$ on the right-hand side ranges over all vertices that have an incident edge contained in $E'$. Because $\sigma_I=\pm 1$, a subset $E'$ contributes to $\mathcal{Z}$ if and only if every vertex in $E'$ has an even degree $d_I(E')$. In essence, this selective elimination acts as a global topological constraint that filters out all open string configurations. Euler's theorem further guarantees that the only surviving edge configurations are closed loops $\Gamma$, thereby producing the loop-gas expansion of the primary theory:
\begin{equation}
\mathcal{Z} \sim \sum_{\Gamma} \prod_{e \in \Gamma} \tanh K_e\,.
\end{equation}
The exact correspondence \eqref{eq:kw_longform} between the couplings themselves then follows immediately by identifying this high-temperature loop-gas expansion with the low-temperature domain-wall expansion of the dual theory. In the isotropic limit \(K_{e} = K\) for all edges \(e\), solving \eqref{eq:kw_longform} for self-duality yields the critical coupling \(K_c = \tfrac{1}{2}\log(1+\sqrt{2})\) at which the system undergoes a phase transition from ferromagnetic order to disorder. At this coupling, duality is a symmetry of the theory and the theory admits a continuum limit described by the Ising CFT.
\subsection{Defect Insertion}
\begin{figure}[]
\centering
\includegraphics[width=0.7\linewidth]{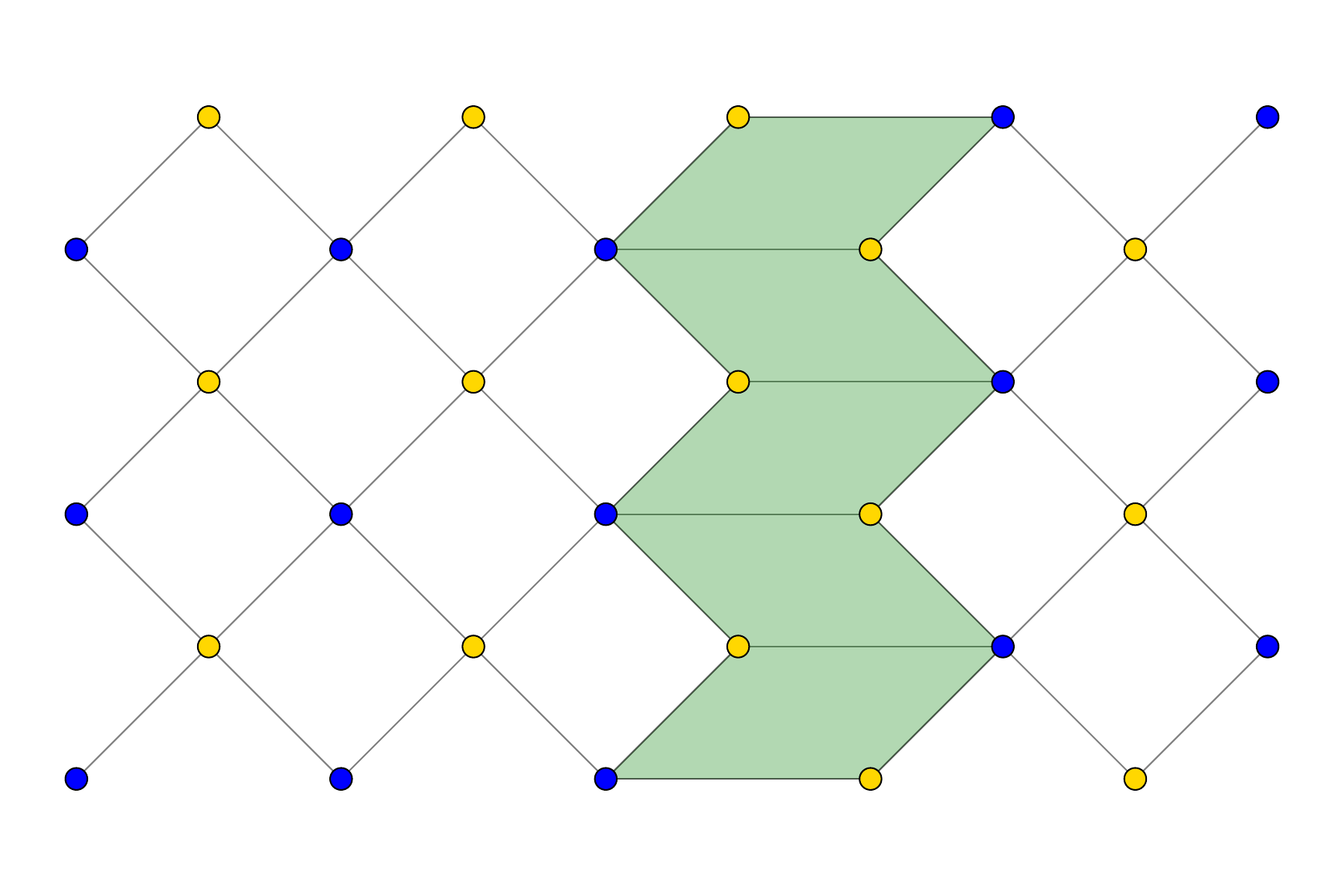}
\caption{Lattice realization of the non-invertible Kramers-Wannier duality defect line. The seam acts as a topological interface matching the primary Hilbert space $\mathcal{H}$ (blue sites) with the dual Hilbert space $\hat{\mathcal{H}}$ (yellow sites), enforcing the non-local duality-twisted boundary conditions across the cycle.}
\label{fig:duality_defect}
\end{figure}

The discrete $\mathbb{Z}_2$ spin-flip symmetry and the Kramers-Wannier duality at the critical point $K_c$ constitute global symmetries of the bulk theory, acting uniformly across the entire lattice configuration space. A central problem in the study of non-local boundary conditions is whether these global actions can be consistently gauged to operate locally. Kramers-Wannier duality being reliant on global identity definitions and sign trivializations is particularly constrained, especially when the underlying lattice has non-trivial topology.
Nonetheless gauging these global symmetries, if possible, contains rich new physics as it would permit the introduction of topological defect lines along the non-contractible $a$ and $b$ cycles of the toroidal lattice \cite{Oshikawa:1996dj}. These defect lines may be realized geometrically as follows. The lattice manifold is partitioned into two distinct spatial domains: a primary region $A$, governed by the original spin variables $\sigma$, and a dual region $B$, formulated in terms of the dual disorder operators $\mu$. As Figure~\ref{fig:duality_defect}, the duality defect line is then defined as the interface or locus across which the non-local Kramers-Wannier change of variables is explicitly executed. If the underlying base manifold possesses a non-trivial topology, the admissibility and consistency of such localized interface insertions become highly non-trivial. 
\begin{figure}[]
\centering
\includegraphics[width=0.7\linewidth]{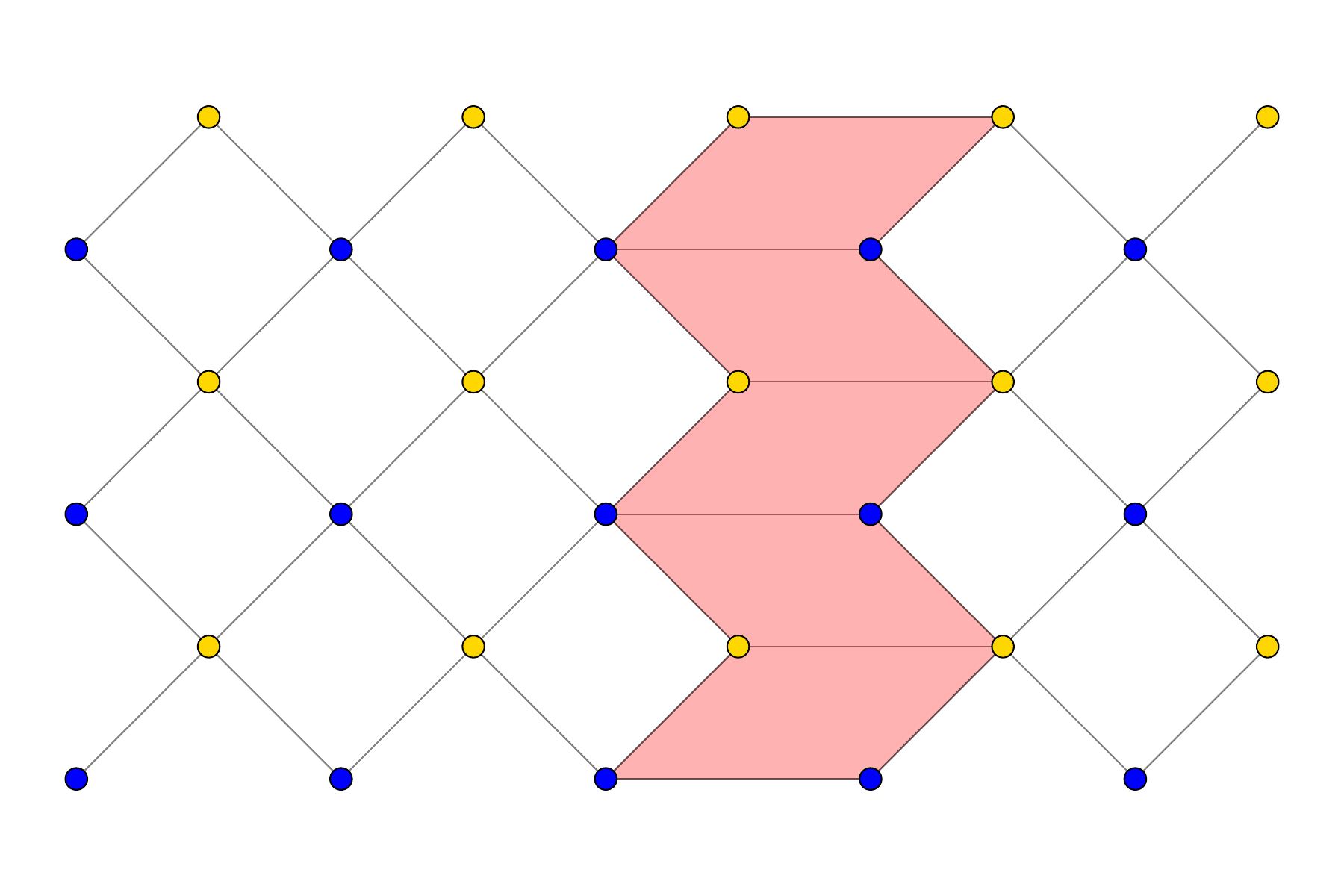}
\caption{Lattice configuration of a vertical $\mathbb{Z}_2$ spin-flip defect line. The defect is implemented by executing a local gauge transformation that systematically inverts the signs of the vertical bond couplings ($K_y \mapsto -K_y$) along the highlighted red 
cut-line switching ferromagnetic interactions across the bond to anti-ferromagnetic. Blue dots represent the primary lattice sites and yellow dots are the dual lattice sites}
\label{fig:spinflip_defect}
\end{figure}

We next review the construction of defect lines configured as closed loops that wind around the non-contractible $a$ and $b$ cycles of the torus background.
A spin-flip defect line is implemented by executing a local $\mathbb{Z}_2$ gauge transformation on a subset of edges $\{\varepsilon\}$:
\begin{equation}
\label{eq:z2gauge}
K_{IJ} \mapsto -K_{IJ} \,, \quad \forall\quad \, \langle IJ \rangle \in \{\varepsilon\}\,.
\end{equation}
A horizontal defect line is constructed by choosing $\{\varepsilon\}$ to be the set of vertical edges $\{e_{\langle (x, y_0), (x, y_0+1) \rangle}\}$ for a fixed row $y_0$, where the horizontal coordinate $x$ runs from $1$ to $L$. This path of modified bonds traces out a non-contractible cycle wrapping around the columns of the torus. Conversely, a vertical defect line corresponds to flipping the horizontal bonds along a fixed column $x_0$, namely $\{e_{\langle (x_0, y), (x_0+1, y) \rangle}\}$, tracing out a cycle wrapping around the $r$ rows of the torus. Inserting these line defects into a reference lattice with periodic-periodic boundary conditions is equivalent to imposing anti-periodic boundary conditions along the corresponding cycles. We anticipate the continuum limit of the lattice theory at criticality and express the corresponding conformal field theory (CFT) partition functions using the defect notation $\mathcal{Z}^{\mathcal{L}_3}_{\mathcal{L}_1 \mathcal{L}_2}$. Here, the lower indices $\mathcal{L}_1$ and $\mathcal{L}_2$ denote the topological defect lines (TDLs) inserted along the spatial $x$-cycle and temporal $y$-cycle, respectively, while the upper index $\mathcal{L}_3$ specifies the allowed fusion channel of the intersecting defects. Under this convention, the relevant sectors are denoted by $\mathcal{Z}^\psi_{\psi\,1}$, $\mathcal{Z}^\psi_{1\psi}$, and $\mathcal{Z}^1_{\psi\psi}$. This notation is summarized in Table~\ref{tab:notation_conventions} along with other principal conventions adopted in the paper. For instance, $\mathcal{Z}^\psi_{\psi\,1}$ represents the CFT partition function with a $\psi$-defect insertion (yielding $\psi$-twisted boundary conditions) along the spatial $x$-cycle, an untwisted identity line along the temporal $y$-cycle, and a net topological flux in the $\psi$ channel.

We show the lattice configuration of a vertical $\mathbb{Z}_2$ spin-flip defect line in Figure~\ref{fig:spinflip_defect}. The physical consequences of a spin-flip defect insertion along such cycles can be tracked by means of certain measureables which we now define. For one, define the following $x$-seam and $y$-seam correlators:
\begin{equation}
\label{eq:seamOps}
O_x = \frac{1}{L}\sum_{i=0}^{L-1} \sigma_{i,L-1}\,\sigma_{i,0}\,, \quad O_y = \frac{1}{L}\sum_{j=0}^{L-1} \sigma_{L-1,j}\,\sigma_{0,j}\,.
\end{equation}
These seam operators track the physical consequences of inserting topological defects by sampling pairs of adjacent spins residing on immediate opposite sides of the chosen boundary cycle. In the untwisted bulk theory, these reduce to standard nearest-neighbor spin-spin correlation functions. However, the insertion of a frustative $\mathbb{Z}_2$ spin-flip defect acts to anti-correlate the spins across the cut. This anti-correlation effect becomes highly pronounced at low temperatures ($K \to \infty$) and vanishes in the high-temperature entropic limit ($K \to 0$), exhibiting a characteristic universal scaling behavior at the conformal critical point. The presence of the spin-flip defect also affects the free energy measurably as it explicitly alters the system's energetics. The modified Hamiltonian reads:
\begin{equation} \label{eq:Hpsi}
H_{\text{defect}} = H_{\mathrm{Ising}} - 2 \sum_{\langle IJ \rangle \in \{\varepsilon\}} K_{IJ} \sigma_I \sigma_J \,,
\end{equation}
which shifts the dimensionless free energy by:
\begin{equation}
\label{eq:dFspinflip}
\Delta F = -\ln \left( \frac{\mathcal{Z}_{\mathrm{defect}}}{\mathcal{Z}_{\mathrm{Ising}}} \right) \,.
\end{equation}
The asymptotic behavior of $\Delta F$ in the thermodynamic limit can be extracted straightforwardly. In the low-temperature (large-$K$) limit, the system develops long-range ferromagnetic order. The spin-flip defect forces neighboring spins across $\{\varepsilon\}$ to anti-align, generating a macroscopic domain wall. The associated energetic penalty scales linearly with the cycle length $L_c$ cut by the seam:
\begin{equation}
\Delta F \approx 2 K L_c \,,
\end{equation}
where $L_c \in \{L, M\}$. In the high-temperature ($K \to 0$) limit, thermal fluctuations dominate, the correlation length vanishes, and the local alignment penalty across the seam becomes negligible compared to the bulk entropy, yielding $\Delta F \to 0$. At the critical point, $\Delta F$ becomes invariant under overall lattice scale variations, governed entirely by universal conformal data.

Inserting a duality defect is, unsurprisingly, more complicated. We introduce a $\mathbb{Z}_2$-valued background gauge field $\eta_e = \pm 1$ on the lattice links \cite{kogut1979}. We configure $\eta_e = -1$ strictly on links crossed by the defect line and $\eta_e = 1$ elsewhere. Under this background, the local flatness constraint generalizes to
\begin{equation}
\prod_{e\in\partial p} \tau_e = \prod_{e\in\partial p} \eta_e\,.
\end{equation}
Away from the defect path, $\prod_{\partial p}\eta_e = 1$ and the conventional flatness condition is preserved. Conversely, any plaquette pierced exactly once by the defect line exhibits a frustrated background $\prod_{\partial p}\eta_e = -1$, forcing a localized $\mathbb{Z}_2$ flux:
\begin{equation}
\prod_{\partial p}\tau_e = -1\,.
\end{equation}

In the absence of defects, the unconstrained substitution $\tau_e = \mu_r\mu_{r'}$ identically satisfies the flatness condition because the product around any closed plaquette evaluates to $\prod_{e \in \partial p}(\mu_r \mu_{r'}) = 1$. Consequently, no globally well-defined dual spin field can reproduce a frustrated plaquette where $\prod_{\partial p}\tau_e = -1$. The defect line thus represents a geometric obstruction to extending the dual spins globally across the manifold. 

This obstruction is formally accommodated via a branch-cut description. The modified constraint is solved by introducing the gauge field directly into the pair-wise mapping:
\begin{equation}
\tau_e = \eta_e \mu_r\mu_{r'} = 
\begin{cases}
\mu_r\mu_{r'}\,, & e \text{ not crossed by the defect}\,, \\
-\mu_r\mu_{r'}\,, & e \text{ crossed by the defect}\,.
\end{cases}
\end{equation}
Taking the product around a plaquette yields $\prod_{\partial p}\tau_e = \prod_{\partial p}\eta_e$, as required. The defect line therefore functions as a branch cut for the dual spins. The global Kramers-Wannier transformation is a non-local map from the primary configurations $\sigma_I$ to dual variables $\mu_{\tilde{I}}$. The defect interface is mediated by the above edge-pair-wise local map above.

Equivalently, the duality transformation can be cast in an operator formalism by mapping the primary site variables $(x_I,y_I)$ to a local state space. Representing the spin eigenbasis $\sigma_I = 1$ and $\sigma_I = -1$ by the orthonormal computational basis vectors $\vert 0\rangle$ and $\vert 1\rangle$ respectively, the global duality map acts as a lattice-wide discrete Fourier transform. Locally, this operation rotates the primary spin Hilbert space $\mathcal{H}_I$ into the dual Hilbert space $\tilde{\mathcal{H}}_{\tilde{I}}$ via the mapping:
\begin{equation}
\vert 0\rangle \mapsto \frac{1}{\sqrt{2}} \left(\vert \tilde{0}\rangle + \vert \tilde{1}\rangle\right)\,, \quad 
\vert 1\rangle \mapsto \frac{1}{\sqrt{2}} \left(\vert \tilde{0}\rangle - \vert \tilde{1}\rangle\right)\,.
\end{equation}
This linear transformation is implemented by the unitary Hadamard operator:
\begin{equation}
\mathrm{H} = \frac{1}{\sqrt{2}} \begin{pmatrix} 1 & 1 \\ 1 & -1 \end{pmatrix}\,.
\end{equation}
In the algebraic framework of anyonic systems and topological quantum field theories (TQFTs), this local basis rotation corresponds precisely to a specific component of the associator or $F$-move matrix, identified here as the fusion $F$-symbol $F^{\sigma}_{\sigma\sigma\sigma}$. 

When localized to a cycle on the torus, the duality defect seam functions as a physical interface that explicitly maps the degrees of freedom from the primary spin Hilbert space $\mathcal{H}$ to the dual space $\tilde{\mathcal{H}}$. This restriction enforces non-local, duality-twisted boundary conditions across the chosen topological cycle. Consequently, the modified Boltzmann weights assigned to the links intersecting this duality defect seam are evaluated as matrix elements of the interface operator:
\begin{equation}
\label{eq:duality_defect_weights}
\mathcal{W}_{\text{defect}}\left(\sigma_I, \mu_{\tilde{J}}\right) = \langle \mu_{\tilde{J}} \vert \mathrm{H} \vert \sigma_I \rangle = \frac{1}{\sqrt{2}} (-1)^{\left(\frac{1-\sigma_I}{2}\right)\left(\frac{1-\mu_{\tilde{J}}}{2}\right)}\,.
\end{equation}
\subsection{Transfer Matrix Expressions} 
The transfer matrix formalism for torus partition functions---both with and without the aforementioned defect insertions---was originally developed in \cite{Aasen:2016dop, Aasen:2020jwb}. Here, we provide a brief review of these expressions, which are comprehensively covered in \cite{Koide:2024auw, Tan:2022vaz}.
To fix conventions, let $V_{p,q}$ denote the vector space corresponding to the spin degrees of freedom at site $(p, q)$. The vector space associated with an entire row $q$ is then given by the tensor product:
\begin{equation}
V_q = \bigotimes_{p=1}^L V_{p,q}\,.
\end{equation}
The row-to-row transfer matrix is a map $T_q: V_q \to V_{q+1}$ defined as:
\begin{equation}
T_q \equiv T^H T^V\,,
\end{equation}
such that the defect-free partition function on a lattice of size $L \times M$ is given by
\begin{equation}
\cZ^1_{1,1}(L,M) = \frac{1}{2^{LM/2}} \tr \left[\left(T^H T^V\right)^M \right]\,.
\end{equation}
Here, $\mathcal{Z}^1_{1,1}$ represents the standard untwisted modular invariant partition function where all insertions are trivial identity lines.

The construction is formulated in terms of the operators:
\begin{equation}
W_p^V = \cos u_V \bI_p + \sin u_V \sigma_p^x\,, \quad 
W^H_{p+1/2} = \cos u_H \bI + \sin u_H \sigma^z_p \otimes \sigma^z_{p+1}\,.
\end{equation}
Here, the operators $W_p^V$ encode interactions along the vertical edges, while $W^H_{p+1/2}$ encodes interactions along the horizontal edges connecting sites $(p,q)$ and $(p+1,q)$ on row $q$. Subscripts denote the specific site spaces upon which these operators act. 
The variables \(u_H\) and \(u_V\) in the above expression are related to the usual Ising couplings \(K_x\) and \(K_y\) via
\begin{equation}\label{eq:transfervar_uv}
\begin{split}
\cos u_V = \frac{e^{K_y}}{\sqrt{e^{2K_y}+e^{-2K_y}}}\,, &\quad \sin u_V = \frac{e^{-K_y}}{\sqrt{e^{2K_y}+e^{-2K_y}}}\,,\\
\cos u_H = \frac{e^{K_x}+e^{-K_x}}{\sqrt{e^{2K_x}+e^{-2K_x}}}\,, &\quad 
\sin u_H = \frac{e^{K_x}-e^{-K_x}}{\sqrt{e^{2K_x}+e^{-2K_x}}}\,.
\end{split}
\end{equation}
The possible site dependence is omitted for brevity. The normalization rescales the Boltzmann partition function so as to make it scale with unit weight under the KW transform. 
These defect insertions are summarized in Table \ref{tab:defects_lines}.
\begin{table}[h!]
\centering
\scriptsize
\renewcommand{\arraystretch}{1.2}
\setlength{\tabcolsep}{3pt}
\begin{tabular}{|c|c|c|}
\hline
\textbf{Defect Insertion} & \textbf{Partition Function} & \textbf{Transfer Matrix Operator(s)} \\ 
\hline
None (periodic) & 
$\cZ_{1,1}(L,M) = \frac{1}{2^{LM/2}} \tr \left[ \left(T^H T^V\right)^M \right]$ & 
$
T^H = \prod_{p=1}^L W^H_{p+1/2} 
T^V = \prod_{p=1}^L W^V_p$ \\
\hline
Horizontal spin-flip & 
$\begin{aligned}
\cZ_{1,\psi}(L,M) &= \tr \left[ \left(T^H T^V\right)^{M-r}\times \right.\\
&\left.\mathcal{D}_\psi \left(T^H T^V\right)^r \right]
\end{aligned}
$ & 
$\mathcal{D}_\psi = \prod_{j=1}^L \sigma_j^x$ \\
\hline
Vertical spin-flip & 
$\cZ_{\psi,1}(L,M) = \tr \left( T^V T_\psi^H \right)^M$ & 
$\begin{aligned}
T_\psi^H &= \left( \cos u_H \bI - \sin u_H \sigma_L^z \sigma_1^z \right) \\
&\quad \times \prod_{j=1}^{L-1} W_{j+1/2}^H
\end{aligned}$ \\
\hline
Horizontal duality & 
$\cZ_{1,\sigma}(L,M) = \tr \left[ \left(T^H T^V\right)^{M-r} \mathcal{D}_\sigma \left(T^H T^V\right)^r \right]$ & 
$\begin{aligned}
\langle \{\hat{h}\} | \mathcal{D}_\sigma | \{h\} \rangle 
&= 2^{-L/2} \prod_{j=1}^L \sigma_j (\hat{\sigma}_{j-1/2} + \hat{\sigma}_{j+1/2})
\end{aligned}$ \\
\hline
Vertical duality & 
$\cZ_{\sigma,1}(L,M) = \tr \left( T_\sigma^V T_\sigma^H \right)^M$ & 
$\begin{aligned}
T_\sigma^V &= \left[ \cos u_V \bI + \sin u_V \sigma_L^z \otimes \sigma_1^x \right] \prod_{j=2}^{L} W_j^V \\
T_\sigma^H &= \prod_{j=1}^{L-1} W_{j+1/2}^H
\end{aligned}$ \\
\hline
\end{tabular}
\caption{Defect insertions in the 2D Ising model using Transfer matrices \cite{Aasen:2016dop}.}
\label{tab:defects_lines}
\end{table}
We also note that these transfer matrices can be used to yield the Boltzmann factors associated with insertions of defect seams. 

\subsection{Monte-Carlo Simulations}
\label{sec:mcsim}
\begin{figure}
    \centering
    \includegraphics[width=0.7\linewidth]{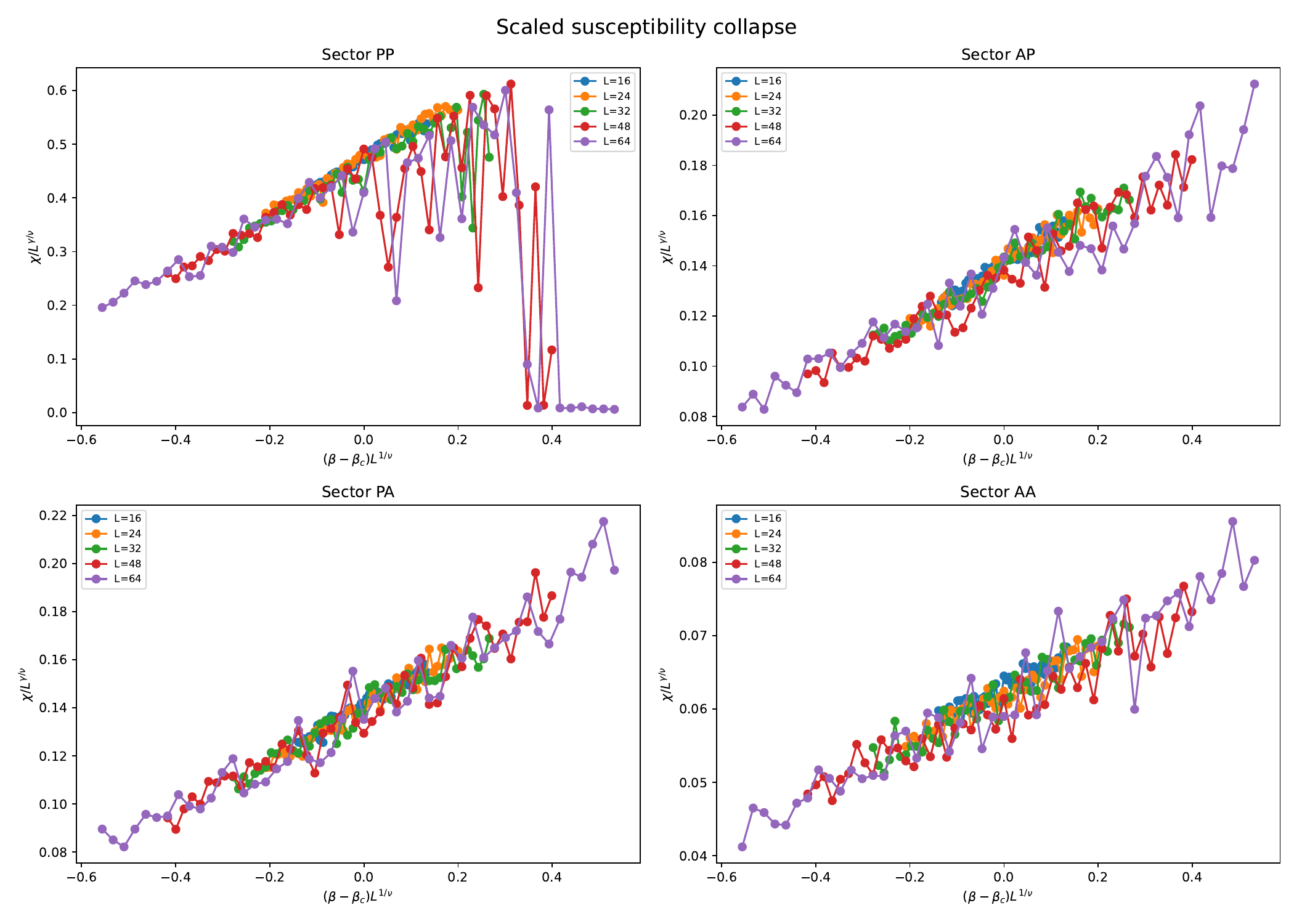}
    \caption{Rescaled susceptibility computed near criticality for the Ising model at varying \(\beta, L\). The plots coalesce over each other, a feature of scale invariance near criticality.}
    \label{fig:chispinflip}
\end{figure}

Thermodynamic properties of the Ising model at thermal equilibrium are encoded in the expectation values of physical observables $\mathcal{O}(\mathbf{s})$ evaluated against the canonical partition function \eqref{eq:ising_partition}. Let $\mathbf{s}$ denote a specific spin microstate on the lattice, defined as the configuration
\begin{equation}
\mathbf{s} = \{\sigma_{I} \quad \forall \quad I\in\Lambda\}\,,
\end{equation}
and let $\mathcal{S}$ denote the configuration space of all such microstates. The thermal expectation value of an observable is given by
\begin{equation}
\left\langle\mathcal{O}\right\rangle = \frac{1}{\mathcal{Z}} \sum_{\mathbf{s}\in\mathcal{S}} \mathcal{O}\left(\mathbf{s}\right)\,e^{-K\,H\left(\mathbf{s}\right)}\,.
\end{equation}
For instance, long-range magnetic order is captured by the microstate magnetization $m(\mathbf{s})$ and its absolute thermal expectation value:
\begin{equation}
m\left(\mathbf{s}\right) = \frac{1}{L^2}\sum_{I\in\Lambda} \sigma_I\,, \quad \left\langle\left\vert m\right\vert\right\rangle = \frac{1}{\mathcal{Z}}\sum_{\mathbf{s}\in\mathcal{S}} \left\vert m\left(\mathbf{s}\right)\right\vert e^{-K\,H\left(\mathbf{s}\right)}\,.
\end{equation}
\begin{figure}
\centering
\includegraphics[width=0.9\linewidth]{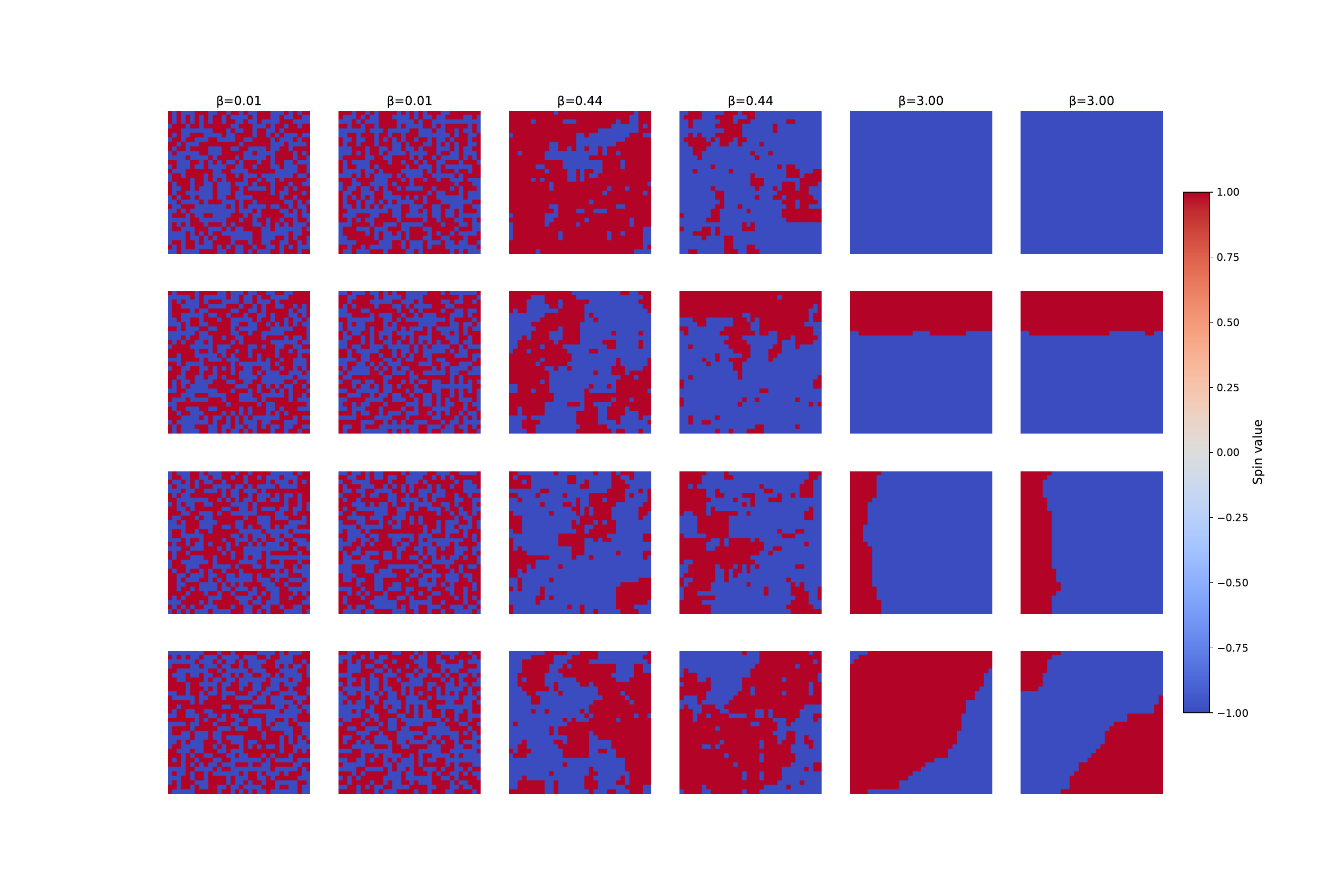}
\caption{Monte Carlo samples of the Ising model on a $64\times64$ lattice, collected from the four spin-flip defect sectors corresponding to
\(\mathcal{Z}^1_{1,1}\), \(\mathcal{Z}^\psi_{1,\psi}\), \(\mathcal{Z}^\psi_{\psi,1}\) and \(\mathcal{Z}^1_{\psi,\psi}\), where we are reading the rows top to bottom . Higher temperatures are shown in columns to the left. The snapshots provide some visual indication of how the antiferromagnetic seam insertion, dominant at low temperatures is negligible in the high temperature limit.}
\label{fig:mcsnapshots}
\end{figure}
Taking the absolute value mitigates catastrophic cancellations in finite systems due to global sign-flip symmetry ($\sigma_I \to -\sigma_I$), while the prefactor $1/L^2$ normalizes the observable against runaway growth originating from the volume of the lattice geometry. Higher-order moments of this distribution yield the magnetic susceptibility:
\begin{equation}
\chi\left(K\right) = L^2 K \left(\left\langle m^2\right\rangle -\left\langle m\right\rangle^2\right)\,.
\end{equation}
We utilize this formal framework to monitor and compare the thermodynamic properties that characterize the spin system both with and without explicit defect insertions. To accurately locate the critical point from numerical simulation data, we evaluate the dimensionless Binder cumulant $U_4(m)$ \cite{binder1, binder2, binder2010monte}:
\begin{equation}
U_4(m) = 1 - \frac{\left\langle m^4\right\rangle}{3\left\langle m^2\right\rangle^2}\,.
\end{equation}
In the thermodynamic limit ($L \to \infty$), $U_4 \to 0$ in the completely disordered high-temperature phase ($K < K_c$), $U_4 \to 2/3$ in the ordered low-temperature phase ($K > K_c$), and at the critical point it takes a universal scale-invariant value $U_4(K_c) \approx 0.61069$ \cite{G_Kamieniarz_1993}. This scale-invariant crossing behavior across different system sizes provides a high-precision baseline for extracting the critical coupling.
To complement these bulk observables, we introduce non-local operators sensitive to the presence of an internal boundary interface. 
In practice one often resorts to estimating these expectation values using Monte-Carlo sampling to draw a sufficiently large number of statistically independent microstates \(\mathbf{s}\) and approximating the above expectation value empirically, i.e. if \(\Omega\) denote the set of independent microstates thus found, we approximate
\begin{equation}
\left\langle\mathcal{O}\left(\mathbf{s}\right)\right\rangle \approx \frac{1}{\vert\Omega\vert}
\sum_{\mathbf{s}\in\Omega} \mathcal{O}\left(\mathbf{s}\right)\,.
\end{equation}
As an example, the rescaled susceptibility near criticality is shown in Figure \ref{fig:chispinflip}.
We have also verified the generated samples against expected Finite Size Scaling behavior.
\section{Wavefunctions of Ising TQFT}
\label{sec:IsingTQFT}
The Ising TQFT is an low-energy effective theory describing non-abelian anyons of Ising type in fractional quantum Hall effect \cite{Willett:1987zz}.  
It contains three anyons $1$, $\psi$ and $\sigma$ with the following non-trivial fusion rules
\[
\psi\times\psi=1,\qquad 
\psi\times\sigma=\sigma,\qquad 
\sigma\times\sigma=1+\psi .
\]
On a torus, the Ising TQFT has a three-dimensional ground-state Hilbert space.
One can choose a basis of these ground states by the anyon flux through a non-contractible cycle. Let $\{|1_x\rangle,\ |\psi_x\rangle,\ |\sigma_x\rangle\}$ be a basis diagonalizing $\sigma$ Wilson loop along $x$-cycle, and let
$
\{|1_y\rangle,\ |\psi_y\rangle,\ |\sigma_y\rangle\}
$
be the corresponding basis for the $y$-cycle. These two bases are related by the modular $S$-matrix,
\begin{equation} \label{eq:isingS}
    \begin{pmatrix}
|1_y\rangle \\[1mm]
|\psi_y\rangle \\[1mm]
|\sigma_y\rangle
\end{pmatrix}
=
S
\begin{pmatrix}
|1_x\rangle \\[1mm]
|\psi_x\rangle \\[1mm]
|\sigma_x\rangle
\end{pmatrix},\quad  S
=
\frac12
\begin{pmatrix}
1&1&\sqrt2\\
1&1&-\sqrt2\\
\sqrt2&-\sqrt2&0
\end{pmatrix}.
\end{equation}
The matrix elements of \(S\) may then be read off from the overlap coefficients 
\begin{equation}
S_{ba} = \langle \Psi^y_b | \Psi^x_a \rangle\, \quad a,b=1,2,3\,.
\end{equation}
with $\Psi^x_a=\{ |1_x\rangle,|\psi_x\rangle,
|\sigma_x\rangle\}$ and $\Psi^y_b=\{ |1_y\rangle,|\psi_y\rangle,|\sigma_y\rangle\}$.
Thus the threefold ground-state degeneracy on the torus, together with the nontrivial relation between the $x$- and $y$-cycle bases, directly realizes the modular data of the Ising TQFT.

\subsection{Proposal of wavefunction from Ising model}
Following the work of \cite{Oshikawa_2007} the Ising model at criticality arises at the boundary of a 3d TQFT with ground state wavefunctions given by
\begin{equation}
\label{eq:FtoIsingx}
    \begin{aligned}
    |\Psi_0\rangle &= \frac{1}{\sqrt{2}}(|1_x\rangle + |\psi_x\rangle) \\
  \mathcal  T_x |\Psi_0\rangle &= \frac{1}{\sqrt{2}}(|1_x\rangle - |\psi_x\rangle) \\
  \mathcal  T_y |\Psi_0\rangle &= |\sigma_x\rangle,
\end{aligned}
\end{equation}
where the tunneling operators $\mathcal T_x$ and $\mathcal T_y$ correspond to creating a pair of $\sigma$-particles and subsequently transporting one of them around the $x$ or $y$ direction and finally annihilating the two. In order to connect this picture to our setup, we first have to map the Ising partition functions at criticality to the corresponding CFT expressions. In the CFT limit, the partition function has the following expression in terms of characters:
\begin{equation}
    Z = |\chi_1|^2 + |\chi_\psi|^2+ |\chi_\sigma|^2\,.
\end{equation}
In terms of spin structures or Majorana fermion boundary conditions, the same partition function is given by
\begin{equation}
    Z = \frac{1}{2}\left(Z_{pp} + Z_{pa} + Z_{ap} + Z_{aa}\right),
\end{equation}
where ``p" and ``a" stand for \textit{periodic} or \textit{antiperiodic} boundary conditions along the corresponding cycles in $Z_{\textrm{space},\textrm{time}}$. The two expressions can be connected once one realizes that the trace over the Neveu-Schwarz (NS) sector gives antiperiodic and the one over the Ramond (R) sector gives periodic boundary conditions in the time directions, while insertion of $(-1)^F$ flips antiperiodic boundary conditions along the spacial cycle to period ones:
\begin{equation} \label{eq:FwithChi}
    \begin{aligned}
Z_{aa} &= \Tr_{\textrm{NS}} \left(q^{L_0 - c/24}\bar q^{\bar L_0 -c/24}\right) = |\chi_1 + \chi_\psi|^2 \\
Z_{ap} &= \Tr_{\textrm{NS}}\left((-1)^{F + \bar F} q^{L_0 - c/24} \bar q^{\bar L_0 - c/24}\right) = |\chi_1 - \chi_\psi|^2 \\
Z_{pa} &= \Tr_{\textrm{R}}\left(q^{L_0-c/24} \bar q^{\bar L_0 -c/24}\right) = 2|\chi_\sigma|^2 \\
Z_{pp} &= \Tr_{\textrm{R}}\left((-1)^{F + \bar F} q^{L_0 - c/24} \bar q^{\bar L_0 - c/24}\right) = 0.
\end{aligned}
\end{equation}
We thus see that taking appropriate ``square roots" of our partition functions produces the wavefunction expressions above, namely 
\begin{align}
\label{eq:squarerootZheuristic}
\sqrt{Z_{aa}} &\sim \frac{1}{\sqrt{2}}(|1_x\rangle + |\psi_x\rangle) \\
\sqrt{Z_{ap}} &\sim \frac{1}{\sqrt{2}}(|1_x\rangle - |\psi_x\rangle) \\
\sqrt{Z_{pa}} &\sim |\sigma_x\rangle.
\end{align}
We will soon clarify in what sense the operation of the taking the square root is implemented. 

\subsection{Partition function with spin-flip defect}
\begin{figure}[ht]
    \centering    \includegraphics[width=.25\textwidth]{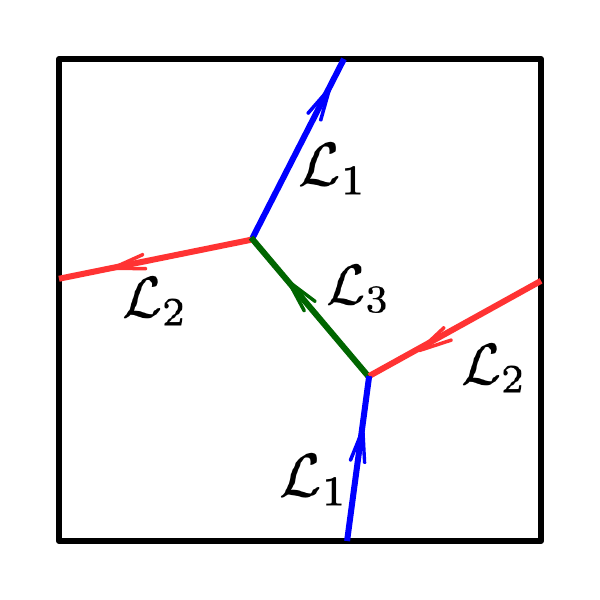}
    \caption{A torus with $\mathcal{L}_1$ inserted in the spatial direction and $\mathcal{L}_2$ inserted in the time direction\,.}
    \label{fig:torustdl}
\end{figure}

In the following we want to rewrite the above expressions in terms topological defect line (TDL) parition functions of the torus. Our notation will be aligned with the diagrammatic depiction in Figure \ref{fig:torustdl} which shows a torus with the TDL $\mathcal{L}_1$ inserted in the spacial direction and $\mathcal{L}_2$ inserted along the time direction. The corresponding partition function is denoted as $Z_{\mathcal{L}_1, \mathcal{L}_2}^{\mathcal{L}_3}$ where $\mathcal{L}_3$ is the fusion channel of the other two defect lines. 
Then, in the case of the Ising model which we study here, the ordinary partition function is given by
\begin{equation}
    Z_{1,1}^1 = \frac{1}{2}\left(Z_{pp} + Z_{pa} + Z_{ap} + Z_{aa}\right)=|\chi_{1}|^2+ |\chi_{\psi}|^2 + |\chi_{\sigma}|^2.
\end{equation}
In order to compute the insertion of a $\psi$ line defect along the spacial or time direction,  one has to compute the resumed periodicity sectors weighted by the \textit{Arf}-invariant \cite{Karch:2019lnn}:
\begin{equation}
    Z_{A_1,A_2}^{A_1+A_2} = \frac{1}{2} \sum_{s_1,s_2} (-1)^{s_1 s_2} Z_{s_1 + A_1, s_2 + A_2},
\end{equation}
where the $A_i$ and the $s_j$ now take values in $\mathbb{Z}_2$ and we use the convention $a = 0$ and $p = 1$ for the boundary conditions. This way one can check the following identities \cite{Petkova:2000ip}:
\begin{align}
Z^\psi_{1,\psi} &= \frac{1}{2} \left( Z_{aa} + Z_{ap} - Z_{pa} + Z_{pp} \right)=|\chi_1|^2
+|\chi_{\psi}|^2-|\chi_{\sigma}|^2, \\
Z^\psi_{\psi,1} &= \frac{1}{2} \left( Z_{aa} - Z_{ap} + Z_{pa} - Z_{pp} \right)= \chi_\psi \chi_1^* + \chi_1 \chi_\psi^*+|\chi_{\sigma}|^2, \\
Z^1_{\psi,\psi} &= \frac{1}{2} \left( -Z_{aa} + Z_{ap} + Z_{pa} + Z_{pp} \right)= -\chi_1{\chi}_{\psi}^* -
\chi_{\psi}{\chi}_1^* + |\chi_{\sigma}|^2.
\end{align}
Partition functions of Majorana fermions can be expressed as   
\begin{equation}\label{eq:FtoIsing}
    \begin{aligned}
        Z_{aa} &= \frac12 \Big( Z_{1,1}^1 + Z^{\psi}_{1,\psi}  + Z^{\psi}_{\psi,1}  - Z^{1}_{\psi,\psi} \Big)=|\chi_1 + \chi_\psi|^2, \\
Z_{ap} &= \frac12 \Big( Z_{1,1}^1 + Z^{\psi}_{1,\psi}  - Z^{\psi}_{\psi,1}  + Z^{1}_{\psi,\psi} \Big)=|\chi_1 - \chi_\psi|^2, \\
Z_{pa} &= \frac12 \Big( Z_{1,1}^1- Z^{\psi}_{1,\psi} + Z^{\psi}_{\psi,1}  + Z^{1}_{\psi,\psi} \Big)=|\chi_{\sigma}|^2, \\
Z_{pp} &= \frac12 \Big( Z_{1,1}^1 - Z^{\psi}_{1,\psi}  - Z^{\psi}_{\psi,1}  - Z^{1}_{\psi,\psi} \Big)=0.
    \end{aligned}
\end{equation}
Since $Z_{pp}=0$, these expressions can be simplified to be  
\begin{equation}\label{eq: ZF2Ising}
\begin{aligned}
Z_{aa} =   Z^{\psi}_{1,\psi}  + Z^{\psi}_{\psi,1} , \quad 
Z_{ap} =   Z_{1,1}^1   - Z^{\psi}_{\psi,1}  , \quad
Z_{pa} =    Z_{1,1}^1- Z^{\psi}_{1,\psi}. 
\end{aligned}
\end{equation}
In this way, we relate partition function of Majorana fermions with those of Ising model.

Let $\tau$ be the complex structure of torus. 
Under modular S-transformation $\tau \to -1/\tau$, the partition functions of different sectors are related by 
\begin{equation}
    Z^{\psi}_{1,\psi} (-1/\tau)=Z^{\psi}_{\psi,1} (\tau), 
    \quad 
    Z^{\psi}_{\psi,1} (-1/\tau)=Z^{\psi}_{1,\psi} (\tau),
\end{equation}
while $Z_{1,1}^1$ and $Z^{1}_{\psi,\psi}$ are invariant. 
From the equation \eqref{eq:FtoIsing}, we have
\begin{equation}
 Z_{aa} (-1/\tau)=Z_{aa} (\tau), \quad     Z_{ap} (-1/\tau)=Z_{pa} (\tau), 
    \quad 
    Z_{pa}(-1/\tau)=Z_{ap} (\tau).
\end{equation}
In terms of matrix as 
\begin{equation} \label{eq:tildeS}
    \begin{pmatrix}
Z_{aa}(-1/\tau) \\[1mm]
Z_{ap}(-1/\tau) \\[1mm]
Z_{pa}(-1/\tau) 
\end{pmatrix}
=
\tilde{S}\begin{pmatrix}
Z_{aa}(\tau) \\[1mm]
Z_{ap}(\tau) \\[1mm]
Z_{pa}(\tau) 
\end{pmatrix}
,\quad 
\tilde{S} = 
\begin{pmatrix}
1 & 0 & 0 \\
0 & 0 & 1 \\
0 & 1 & 0 \\
\end{pmatrix}
\end{equation}
Take ``square root" of these partition functions and transform them back to Ising characters 
with equation \eqref{eq:FwithChi}, we have 
\begin{equation}
    \begin{pmatrix}
 \sqrt{Z_{aa}}(\tau)\\[1mm]
 \sqrt{Z_{ap}}(\tau)\\[1mm]
 \sqrt{Z_{pa}}(\tau)
\end{pmatrix}
= M \begin{pmatrix}
\chi_1(\tau)\\[1mm]
\chi_\psi(\tau)\\[1mm]
\chi_\sigma(\tau)
\end{pmatrix},\quad M=
\begin{pmatrix}
1 & 1 & 0\\
1 & -1 & 0\\
0 & 0 & \sqrt{2}
\end{pmatrix}
\end{equation}
It's straightforward to have 
\begin{equation} \label{eq:SmatrixZ}
    S = M^{-1} \tilde{S} M = \frac{1}{2}
\begin{pmatrix}
1 & 1 & \sqrt{2} \\
1 & 1 & -\sqrt{2} \\
\sqrt{2} & -\sqrt{2} & 0
\end{pmatrix}
\end{equation}
which is the S-matrix of Ising CFT.

\subsection{RBM wavefunctions}
We now need to define the precise operation of taking the square root of the above partition functions in order to arrive at the 3d TQFT boundary wavefunctions. To motivate this mapping, we frame the partition function through the ansatz:
\begin{equation}
\label{eq:state_purification}
\mathcal{Z} = \sum_{\mathbf{v}\in\mathcal{V}} \mathrm{p}\left(\mathbf{v}\right) \equiv \langle \Psi \vert \Psi \rangle\,,
\end{equation}
where $\mathrm{p}(\mathbf{v})$ represents the un-normalized Boltzmann weight of the classical microstate $\mathbf{v}$, and $\vert\Psi\rangle$ is an un-normalized quantum state vector. Under a suitably optimized RBM configuration in Section \ref{sec:IsingRBM}, this classical partition function is expressed up to a global normalization factor as
\begin{equation}
\mathcal{Z}_{\mathrm{RBM}} = \sum_{\mathbf{v}\in\mathcal{V}}\sum_{\mathbf{h}\in\mathcal{H}}  e^{\mathbf{v}^T \mathbf{W} \mathbf{h} + \mathbf{b}^T \mathbf{h} + \mathbf{c}^T \mathbf{v}} \equiv \sum_{\mathbf{v}\in \mathcal{V}} \Phi_{\boldsymbol{\theta}}\left(\mathbf{v}\right)\,,
\end{equation}
where $\boldsymbol{\theta}=\{\mathbf{W},\mathbf{b},\mathbf{c}\}$ denotes the network parameters implicitly optimized to match the target partition function. \(\Phi_{\boldsymbol{\theta}}\left(\mathbf{v}\right)\) is clearly just the marginalized distribution \(\mathrm{p}(\mathbf{v})\), but we separate notation for clarity.
From this configuration space, we define the component-wise square-root operation that maps the classical Boltzmann weights directly to the wavefunction:
\begin{equation}
\label{eq:rmbZwavefunction}
\vert \Psi \rangle = \sum_{\mathbf{v}\in\mathcal{V}} \Psi_{\boldsymbol{\theta}}(\mathbf{v}) \vert \mathbf{v} \rangle = \sum_{\mathbf{v}\in\mathcal{V}} \sqrt{\mathrm{\Phi}_{\boldsymbol{\theta}}(\mathbf{v})}\,\vert\mathbf{v}\rangle\,.
\end{equation}
Here $\vert\mathbf{v}\rangle$ is a microscopic quantum state represented by spin configuration on lattice. Thus, one square lattice with length $L$. The dimension of Hilbert space is $2^{L^2}$.  
To simplify subsequent calculations, we adopt a compact functional shorthand for this state vector mapping, writing $\Psi = \sqrt{\Phi}$, suppressing all other indices. This furnishes the precise definition of the square root operation of \eqref{eq:squarerootZheuristic}.

In particular, the partition functions \(\mathcal{Z}_{ab}^c\) define state vectors \(\Phi^c_{ab}\) as above. From equation \eqref{eq: ZF2Ising}, we can find the state vectors corresponding to $Z_{aa}, Z_{ap}$ and $Z_{pa}$ as 
\begin{equation}
\begin{aligned}
\Phi_{aa} =  \Phi^{\psi}_{1,\psi}  + \Phi^{\psi}_{\psi,1}  ,\quad \Phi_{ap}= \Phi_{1,1}^1   - \Phi^{\psi}_{\psi,1},\quad \Phi_{pa}=\Phi_{1,1}^1- \Phi^{\psi}_{1,\psi}\;.
\end{aligned}
\end{equation}
From equation \eqref{eq:squarerootZheuristic}, we obtain the following wavefunctions 
\begin{equation}
\label{eq:ansatz2}
\begin{aligned}
    |\Psi_0\rangle & \sim   \Psi_{aa}=\sqrt{\Phi_{aa}},\\ 
T_x |\Psi_0\rangle & \sim   \Psi_{ap}=\sqrt{\Phi_{ap}},\\ 
T_y |\Psi_0\rangle &  \sim \Psi_{pa} = \sqrt{ \Phi_{pa}}
\end{aligned}
\end{equation}
By a change of basis \eqref{eq:FtoIsingx}, the basis diagonalizing $W_{\sigma}$ along $x$-cycle is 
\begin{equation} \label{eq:xbasisF}
\begin{pmatrix}
|1_x\rangle \\[1mm]
|\psi_x\rangle \\[1mm]
|\sigma_x\rangle
\end{pmatrix}
=
M^{-1}
\begin{pmatrix}
\Psi_{aa}(\tau)\\[1mm]
 \Psi_{ap}(\tau)\\[1mm]
\Psi_{pa}(\tau)
\end{pmatrix},
\quad 
M^{-1}=
\frac{1}{2}
\begin{pmatrix}
1 & 1 & 0\\
1 & -1 & 0\\
0 & 0 & \sqrt{2}
\end{pmatrix}
\end{equation}
where modular parameter $\tau$ is kept explicit here.

The basis diagonalizing $W_{\sigma}$ along $y$-cycle can be obtained from the basis along x-cycle by S-transformation. 
Consider a rectangle lattice $\Lambda$ of size $L_x \times L_y$. 
The spins on $\Lambda$ are denoted by $\sigma_{i,j}$ with $i\in[0,L_x-1]$ and $j\in[0,L_y-1]$. The S-transformation $\tau \to -1/\tau$ changes $\Lambda$ to a lattice $\hat{\Lambda}$ of size $L_y \times L_x$. The spin configurations $\{\sigma_{ij}\} $ on $\Lambda$ are mapped to those on $\hat{\Lambda}$ by  
\begin{equation} \label{eq:sspin}
   s:\;\; \sigma_{ij} \to \sigma_{-j,i}\;.
\end{equation}
where $(-j) \mod{L_y}$ is taken due to the periodical boundary condition. By S-transformation, we obtain wavefunctions on $\hat{\Lambda}$   
\begin{equation} 
\label{eq:tildeSPsi}
    \Psi_{aa}(-1/\tau) =  \Psi_{aa}(\tau),\quad \Psi_{ap}(-1/\tau) = \Psi_{pa}(\tau),\quad 
    \Psi_{pa}(-1/\tau) = \Psi_{ap}(\tau)\;.
\end{equation}
In terms of matrix, they are related by a matrix $\tilde{S}$ defined in \eqref{eq:tildeS}. 
Note that two wavefunctions defined on lattice $\Lambda$ with modular parameter $\tau$ and  $\hat{\Lambda}$ with $-1/\tau$ are equal if they have the value on spin configuration $\{\sigma_{ij}\}$ on $\Lambda$ and $\{\sigma_{-ji}\}$ on $\hat{\Lambda}$.

Thus, the basis diagonalizing $W_{\sigma}$ along y-cycle can be obtained from equation \eqref{eq:xbasisF} by replacing $\tau \to -1/\tau$ as 
\begin{equation} \label{eq:ybasisF}
\begin{pmatrix}
|1_y\rangle \\[1mm]
|\psi_y\rangle \\[1mm]
|\sigma_y\rangle
\end{pmatrix}
=
M^{-1}
\begin{pmatrix}
\Psi_{aa}(-1/\tau)\\[1mm]
\Psi_{ap}(-1/\tau)\\[1mm]
\Psi_{pa}(-1/\tau)
\end{pmatrix}
\end{equation}
These two basis are related by S-matrix of Ising TQFT. Indeed, from equation \eqref{eq:tildeSPsi} and \eqref{eq:xbasisF}, we can check that  
\begin{equation}
\begin{pmatrix}
|1_y\rangle \\[1mm]
|\psi_y\rangle \\[1mm]
|\sigma_y\rangle
\end{pmatrix}
=
M^{-1}
\begin{pmatrix}
{\Psi}_{aa}(-1/\tau)\\[1mm]
{\Psi}_{ap}(-1/\tau)\\[1mm]
{\Psi}_{pa}(-1/\tau)
\end{pmatrix}
=
M^{-1}\tilde{S}
\begin{pmatrix}
{\Psi}_{aa}(\tau)\\[1mm]
{\Psi}_{ap}(\tau)\\[1mm]
{\Psi}_{pa}(\tau)
\end{pmatrix}
=
M^{-1}\tilde{S} M 
\begin{pmatrix}
|1_x\rangle \\[1mm]
|\psi_x\rangle \\[1mm]
|\sigma_x\rangle
\end{pmatrix}
\end{equation}
Here $M^{-1}\tilde{S}M$ is the S-matrix of Ising TQFT defined in equation \eqref{eq:SmatrixZ}. The inner product between these two basis along is expected to give S-matrix of Ising TQFT \eqref{eq:isingS}. We will check it using RBM wavefunction in next section.  

\section{Representing the Ising Model on RBMs}
\label{sec:IsingRBM}
We present our results on training Restricted Boltzmann Machines on Ising spin microstates at criticality with and without insertion of defects. Microscopically, we know that the Ising model Hamiltonian and state space can be rewritten exactly an RBM. The construction implicitly relies on being able to completely sample the \(2^{L^2}\)-dimensional configuration space and is closer to being a proof of existence. Finding an optimal parametric configuration via empirical training is a fundamentally different problem. Optimization requires interpolating from a sparse training ensemble ($N \ll 2^{L^2}$), navigating the non-convex landscape of the inductive bias of gradient descent, and mitigating sampling noise. As is well known, statistical learners, especially highly parametrized ones, often tend to overfit. i.e. the model tends to memorize the available data at the expense of the true regularities of the underlying probability distribution itself. Nonetheless, it has been observed that the RBM does recover imprints of the underlying Hamiltonian and hence the Boltzmann distribution \cite{Iso:2018yqu,ShibaFunai:2018aaw,Cossu:2018pxj}. The analysis below provides an affirmative answer to the question even for more global properties such as defect insertions and additionally addresses questions of symmetry that were not authored into the architecture.

\subsection{Overview on RBMs}
\begin{figure}
\centering
\includegraphics[width=0.5\linewidth]{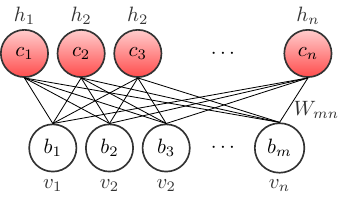}
\caption{RBM Network Graph. The visible nodes \(\mathbf{v}\) are blank and the hidden nodes \(\mathbf{h}\) are shaded red. The parameters \(\mathbf{W}\), \(\mathbf{b}\) and \(\mathbf{c}\) are defined in Equation \eqref{eq:RBMenergy}. }
\label{fig:rbmnetgraph}
\end{figure}

Generative modeling seeks to approximate an in-principle unknown probability distribution \(p\left(\mathbf{v}\right)\) by extrapolating from a finite dataset \(\mathcal{D}\)
\begin{equation}
\mathcal{D} = \left\lbrace \mathbf{v}_i \quad \vert \quad i=1,\,\ldots,\,N\right\rbrace\,,
\end{equation}
containing vectors \(\mathbf{v}\in \mathbb{R}^{n_v}\), presumed to be independent samples drawn from \(p\left(\mathbf{v}\right)\). Having learnt \(p\left(\mathbf{v}\right)\), one could in principle sample from the distribution to generate new data. In the context of systems like the Ising model, it provides the possibility of learning the underlying Boltzmann partition function \(\mathcal{Z}\left(\beta\right)\) from some number of Monte-Carlo samples. In practice, arriving at the true underlying pdf from a finite number of samples, however many, is an intractable problem in much the same way as uniquely determining a smooth curve passing through a finite number of points in \(\mathbb{R}^2\). As in the curve interpolation problem, one requires additional guiding principles and assumptions. For instance, one may hypothesize an ansatz class of functions \(\mathrm{p}_{\theta}\left(\mathbf{v}\right)\) and use the principle of maximum likelihood estimation (MLE) 
\cite{ESLR,ISLR,bengio2017deep}. MLE hypothesizes that the paramters \(\boldsymbol{\theta}\) are determined by maximizing the likelihood of observing the dataset \(\mathcal{D}\). Maximization of the likelihood is equivalent to minimizing the negative log-likelihood, which leads to the objective function
\begin{equation}
\label{eq:mleloss}
\mathcal{L}(\boldsymbol{\theta}) = -\frac{1}{N}\sum_{n=1}^N \log \mathrm{p}_{\boldsymbol{\theta}}(\mathbf{v}_n)\,.
\end{equation}
\(\mathcal{L}\left(\theta\right)\) is a sum of negative logs of arguments between 0 and 1. Hence, it is hence positive semi-definite, vanishing only if the likelihood of observing each datum \(v_n\) would be 1. This formulation is equivalent to minimizing the Kullback-Leibler divergence between the empirical probability distribution \(\mathrm{p}_E\left(\mathbf{v}\right)\) given by
\begin{equation}
\mathrm{p}_E\left(\mathbf{v}\right) = \left\lbrace\begin{matrix}
    1\,&;\quad \text{if } \mathbf{v}\in\mathcal{D}\,,\\
    0\,&;\quad \mathrm{otherwise}\,.
\end{matrix}\right.\,
\end{equation}
and the model probability \(\mathrm{p}_{\boldsymbol{\theta}}\left(\mathbf{v}\right)\).
This estimation problem for the Ising model has therefore also been formulated as a KL divergence minimization problem as in
\cite{Iso:2018yqu,ShibaFunai:2018aaw,funai2021feature}. 
Restricted Boltzmann Machines are an instance of an ansatz family \(\mathrm{p}_{\boldsymbol{\theta}}\left(\mathbf{v}\right)\) which is particularly adapted to the case where \(\mathbf{v}\) is 
binary-valued, i.e. $\mathbf{v} \in \{0,1\}^{n_v}$ (or $\{-1,+1\}^{n_v}$) \cite{smolensky1986information}, see \cite{fischerRBMintro,montufar2016restricted} for reviews. 
The RBM, as shown in Figure \ref{fig:rbmnetgraph}, is defined on a bipartite graph consisting of a visible layer $\mathbf{v}$ and a hidden latent layer $\mathbf{h} \in \{0,1\}^{n_h}$ (or $\{-1,+1\}^{n_h}$). The joint probability distribution of \(\mathbf{v}\) and \(\mathbf{h}\) obeys the Boltzmann distribution:
\begin{equation}\label{eq:jpdf}
\mathrm{p}_{\boldsymbol{\theta}}(\mathbf{v},\mathbf{h}) = \frac{e^{-E(\mathbf{v},\mathbf{h};\boldsymbol{\theta})}}{\mathcal{Z}(\boldsymbol{\theta})}\,,\quad \mathcal{Z}(\boldsymbol{\theta}) = \sum_{\mathbf{v},\mathbf{h}} e^{-E(\mathbf{v},\mathbf{h};\boldsymbol{\theta})}\,.
\end{equation}
The energy function appearing above is constrained by the restriction that no intra-layer connections exist:
\begin{equation}
\label{eq:RBMenergy}
E(\mathbf{v},\mathbf{h};\boldsymbol{\theta}) = -\mathbf{v}^T \mathbf{W} \mathbf{h} - \mathbf{b}^T \mathbf{h} - \mathbf{c}^T \mathbf{v}\,,
\end{equation}
where the parameter set $\boldsymbol{\theta} = \{\mathbf{W}, \mathbf{b}, \mathbf{c}\}$ comprises the weight matrix $\mathbf{W} \in \mathbb{R}^{n_v \times n_h}$ and the bias vectors $\mathbf{b} \in \mathbb{R}^{n_h}$, $\mathbf{c} \in \mathbb{R}^{n_v}$. 
The target probability distribution \(\mathrm{p}\left(\mathbf{v}\right)\) 
in modeled by the function \(\mathrm{p}_{\boldsymbol{\theta}}\left(\mathbf{v}\right)\) 
obtained by marginalizing the RBM function \(\mathrm{p}_{\boldsymbol{\theta}}\left(\mathbf{v}\right)\)
\begin{equation}
\label{eq:rbm_ansatz_ptheta_v}
\log \mathrm{p}_{\boldsymbol{\theta}}(\mathbf{v}) = \log\sum_{\mathbf{h}} e^{-E(\mathbf{v},\mathbf{h};\boldsymbol{\theta})} - \log \mathcal{Z}(\boldsymbol{\theta})\,.
\end{equation}
The RBM ansatz is `restricted' in the sense that the ansatz forbids direct connections within the visible and the hidden nodes themselves. 
Optimization of $\mathcal{L}(\boldsymbol{\theta})$ proceeds via gradient descent \(\theta\) using. We mention a few salient features of the algorithm, referring the reader to \cite{hintoncd} for details.
To begin, note that differentiating \eqref{eq:rbm_ansatz_ptheta_v} with respect to the parameters yields the standard identity:
\begin{equation}
\label{eq:rbmgradients}
\partial_{\boldsymbol{\theta}}\left[-\log \mathrm{p}_{\boldsymbol{\theta}}(\mathbf{v})\right] = \mathbb{E}_{\mathbf{h}\vert \mathbf{v}}\left[\partial_{\boldsymbol{\theta}} E(\mathbf{v},\mathbf{h};\boldsymbol{\theta}) \right] - \mathbb{E}_{\mathbf{v},\mathbf{h}}\left[\partial_{\boldsymbol{\theta}} E(\mathbf{v},\mathbf{h};\boldsymbol{\theta}) \right]\,.
\end{equation}function
The first term, the \textit{positive phase}, can be evaluated analytically (see Appendix \ref{app:cd}):
\begin{equation}
\label{eq:posgradsigmoid}
\partial^+_{\mathbf{W}_{ij}}E = - \mathbf{v}_i{\sigma}(\mathbf{z}_j)\,,\quad
\partial^+_{\mathbf{b}_{j}}E = -{\sigma}(\mathbf{z}_j)\,,\quad
\partial^+_{\mathbf{c}_i}E = -\mathbf{v}_{i}\,,
\end{equation}
where $\mathbf{z} = \mathbf{v}^T \mathbf{W} + \mathbf{b}^T$ and $\sigma(\mathbf{z}) = (1 + e^{-\mathbf{z}})^{-1}$ denotes the element-wise logistic sigmoid function. Conversely, the second term---the \textit{negative phase}---requires calculating an expectation over the full joint distribution $\mathrm{p}_{\boldsymbol{\theta}}(\mathbf{v},\mathbf{h})$. Because evaluating $\mathcal{Z}(\boldsymbol{\theta})$ is generally intractable, this term must be approximated. 
Having computed the gradients as shown, we train the RBM against the loss function defined via \eqref{eq:mleloss} and \eqref{eq:rbm_ansatz_ptheta_v} as one would any other ML algorithm.

Such estimations require methods such as the contrastive divergence ($k$-CD) algorithm \cite{hinton2012,hintoncd}, generating an approximate sample step via a $k$-step Gibbs sampler initiated at the data point $\mathbf{v}^{(0)} \equiv \mathbf{v}$:
\begin{equation}
\mathbf{h}^{(m)} \sim \mathrm{p}(\mathbf{h}\vert \mathbf{v}^{(m-1)})\,,\quad \mathbf{v}^{(m)} \sim \mathrm{p}(\mathbf{v}\vert \mathbf{h}^{(m)})\,,\quad m = 1, \dots, k\,.
\end{equation}
The resulting estimator for the negative phase gradients is given by:
\begin{equation}
\partial^-_{\mathbf{W}_{ij}}E = \mathbf{v}_i^{(k)}\,\mathbf{h}_j^{(k)}\,,\quad 
\partial^-_{\mathbf{b}_j}E = \mathbf{h}_j^{(k)}\,,\quad 
\partial^-_{\mathbf{c}_i}E = \mathbf{v}_i^{(k)}\,.
\end{equation}
Empirically, setting $k=1$ often provides sufficient convergence for reliable optimization \cite{hinton2012,hintoncd}. In this paper we estimate gradients by \emph{persistent contrastive divergence} \cite{tieleman2008training}\,, and train the RBM by stochastic gradient descent while monitoring metrics such as the reconstruction error, free energy and pseudo-likelihood, defined in Appendix \ref{app:RBMtrain}. 
 
\subsection{Latent Encoding of Defect Seams and Ising Interactions}
\begin{figure}[h]
\centering
\includegraphics[height=0.2\textheight, width=0.8\textwidth]{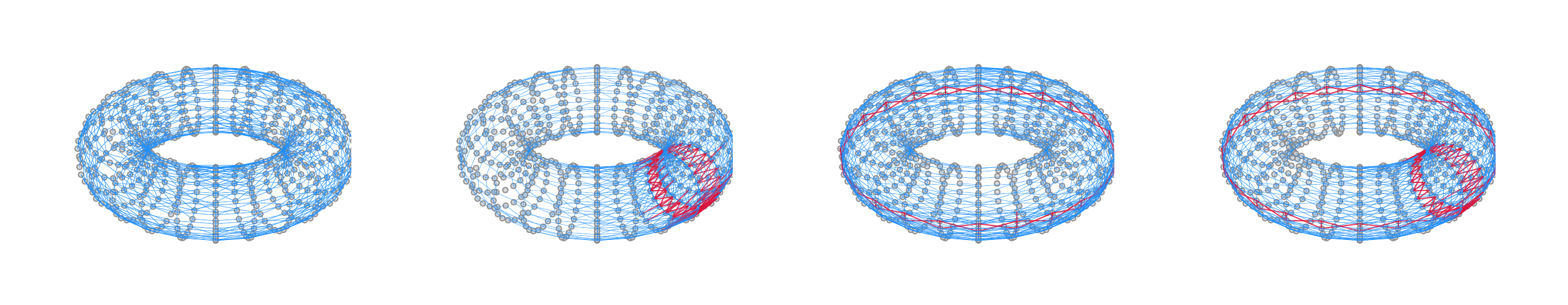}
\caption{Spatial projection of the RBM weight covariance matrix $\Sigma(\mathbf{v}_i,\mathbf{v}_j) = \mathbf{W}\mathbf{W}^T$ trained on snapshots drawn from the $\mathcal{Z}_{1,1}$, $\mathcal{Z}_{1,\psi}$, $\mathcal{Z}_{\psi,1}$, and $\mathcal{Z}_{\psi,\psi}$ sectors. The matrix entries are mapped as links between coordinate sites (blue dots) on a toroidal geometry. Links are colored blue for positive (ferromagnetic) values and red for negative (anti-ferromagnetic) values, providing visual confirmation that the latent layer faithfully recovers the non-contractible spin-flip defect seam.}
\label{fig:rbm_covariance_torus}
\end{figure}

Before examining the empirical results, it is instructive to establish how an RBM mathematically captures the underlying physical interactions of the training data. The joint probability distribution defined in \eqref{eq:jpdf} is conditioned on an energy function $E(\mathbf{v},\mathbf{h};\boldsymbol{\theta})$ where no intra-layer connections exist. To see how visible-to-visible correlations emerge, we can analytically marginalize over the hidden degrees of freedom. For binary states $\mathbf{h} \in \{0, 1\}^{n_h}$, the marginal distribution reads:
\begin{equation}
\label{eq:prbm_visible}
\mathrm{p}_{\boldsymbol{\theta}}(\mathbf{v}) = \frac{e^{\mathbf{c}^T \mathbf{v}}}{\mathcal{Z}(\boldsymbol{\theta})}  \prod_{j=1}^{n_h} 
\left( 1 + e^{\left(\mathbf{v}^T \mathbf{W}\right)_{j} + b_j} \right)\,.
\end{equation}
This representation may quite naturally be identified to an `effective action' for the RBM from which one may extract couplings \cite{Cossu:2018pxj}. It is instructive to recollect some elements of the computation for the generator of two-point functions which we may interpret as a quadratic Hamiltonian over site-space \(H_{ij}\). The energy functional \(\mathrm{E}\left(\mathbf{v}\right)\) corresponding to \eqref{eq:prbm_visible} when expanded in components takes the form
\begin{equation}
\mathrm{E}\left(\{v_1,\,v_2,\,\ldots v_n\}\right) = 
-\sum_{k=1}^{n_v} c_k\,v_k-\sum_{j=1}^{n_h} 
\log\left( 1 + e^{\sum_{k=1}^{n_v} v_k\,W_{kj} + b_j} \right)\,.
\end{equation}
Assuming that the entries of \(\mathbf{W}\) are small compared to \(\mathbf{b}\), series expansion to quadratic order in \(\left(\mathbf{v}\cdot\mathbf{W}\right)_j\) yields the heuristic
\begin{equation}
\label{eq:effective_coupling}
J_{ij}^{\text{eff}} \propto \sum_{k=1}^{n_h} W_{ik}\sigma\left(b_k\right)\sigma\left(b_k\right) W_{kj}^T = \left(\mathbf{W}\mathbf{W}^T\right)_{ij}\,.
\end{equation} 
We empirically verify the parameter estimation across different sectors in Figure 
\begin{figure}
\centering
\includegraphics[width=0.8\linewidth]{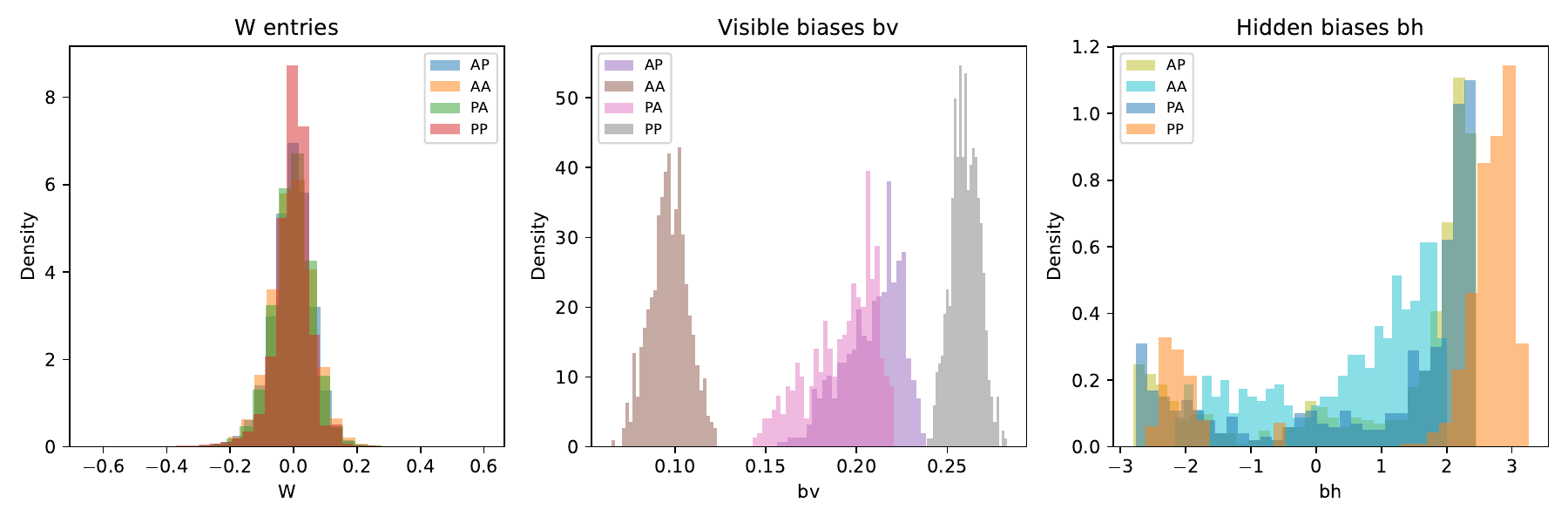}
\caption{Visualizing spread in parameters of the trained RBM. The parameters \(\mathbf{W}\) and \(\mathbf{c}\) have finite spread but are close to zero. The only order one parameters are the \(\mathbf{b}\), argued when writing the perturbative expansion.}
\label{fig:placeholder}
\end{figure}
In particular, we expect that nearest neighbour interactions and other short range effects be mediated by the \(\mathbf{W}\) parameters while the bias vectors mediate emergent, long-range interactions. This suggests that effects which can be traced back to localized deformations would be chiefly encoded within the structural profile of $\mathbf{W}\mathbf{W}^T$. 
\begin{figure}[h]
\centering
\includegraphics[width=\linewidth]{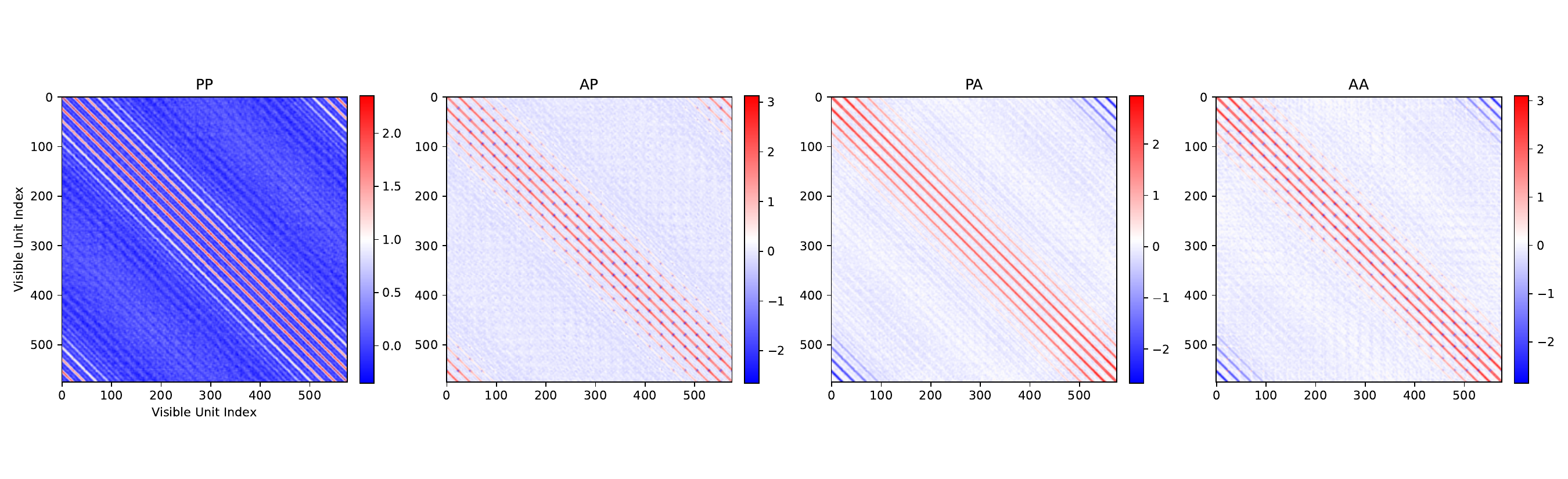}
\caption{The RBM weight covariance matrix $\Sigma(\mathbf{v}_i,\mathbf{v}_j) = \mathbf{W}\mathbf{W}^T$ for distinct topological sectors plotted as heatmaps. The appearance of anti-ferromagnetic couplings is distinctly visible in sectors where topological defects are present.}
\label{fig:rbm_covariance}
\end{figure}
Remarkably, this expectation is precisely borne in Figures \ref{fig:rbm_covariance} and \ref{fig:rbm_covariance_torus} which show how the covariance matrix recovers the insertion of the antiferromagnetic seams even at criticality. 

\begin{figure}
\centering
\includegraphics[width=0.8\linewidth]{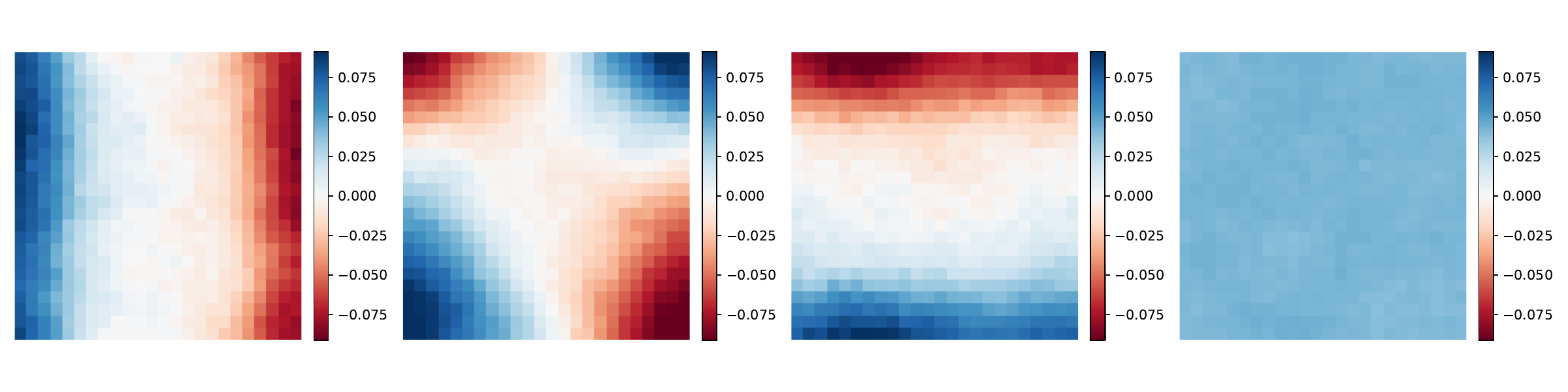}
\caption{Leading eigenmodes obtained by diagonalizing \eqref{eq:effective_coupling} in each sector. The sectors are, left to right, \(\mathcal{Z}^{\psi}_{1\psi},
\mathcal{Z}^{1}_{\psi\psi},\mathcal{Z}^{\psi}_{\psi 1},\mathcal{Z}^{1}_{11}\). The lowest energy mode in \(\mathcal{Z}^1_{11}\) is the zero momentum mode.}
\label{fig:eigenmodes}
\end{figure}

Of course, in an idealized setting, one expects the learned covariance matrix $\mathbf{W}\mathbf{W}^T$ to correspond directly to the microscopic Hamiltonian, which is quadratic in the spin variables. This mapping, however, is significantly complicated by the physical realities of the training dataset: the RBM is trained on snapshots generated exactly at criticality, where the spatial correlation length diverges. In the continuum limit, this divergence effectively manifests as the generation of long-range, non-nearest-neighbor correlation terms in the effective visible energy. Furthermore, finite-size sampling artifacts, risks of overfitting stemming from over-parametrization, and the dense connectivity of the latent layer to the visible nodes inevitably introduce structural noise. Nonetheless, despite these competing factors, the dominant low-energy features captured by the RBM parameters faithfully preserve and encode the underlying geometric and topological correspondence. This is visible for instance in the weight covariance heatmaps shown in Figure \ref{fig:rbm_covariance}. The non-zero entries reside strictly along the main diagonal and specific off-diagonal bands. These bands are a direct artifact of unrolling the 2D spatial lattice into a 1D vector; they denote nearest-neighbor interactions on the torus that were spatially separated by the string serialization. 
Crucially, when spin-flip defects are present in the training ensembles, we observe the appearance of regularly spaced, negative (anti-ferromagnetic) interactions. These localized sign reversals correspond exactly to the coordinates where the primary ferromagnetic bonds were inverted ($K_{IJ} \mapsto -K_{IJ}$) to construct the defect seam. To visualize this spatial alignment intuitively, the learned parameters of $\mathbf{W}\mathbf{W}^T$ can be mapped directly back onto the original 2D toroidal lattice geometry by representing the matrix elements as physical links between site coordinates, as illustrated in Figure \ref{fig:rbm_covariance_torus}.
\paragraph{Torus Ground States from the RBM:}
We next study the low-energy excitations of each sector. These are encoded in eigenmodes corresponding to the leading eigenvalues of \eqref{eq:effective_coupling}. The results obtained for the leading eigenmodes in each sector are shown in Figure \ref{fig:eigenmodes}.
Visualizing these modes generated from the RBM is quite insightful. The lightest mode in each sector is in essence dictated by the choice of defect insertion along the torus. The eigenvalues decay relatively slowly in each sector, which possibly reflects the multiple interaction effects which cannot be parametrically neglected at criticality.
\begin{figure}
\centering
\includegraphics[width=0.4\linewidth]{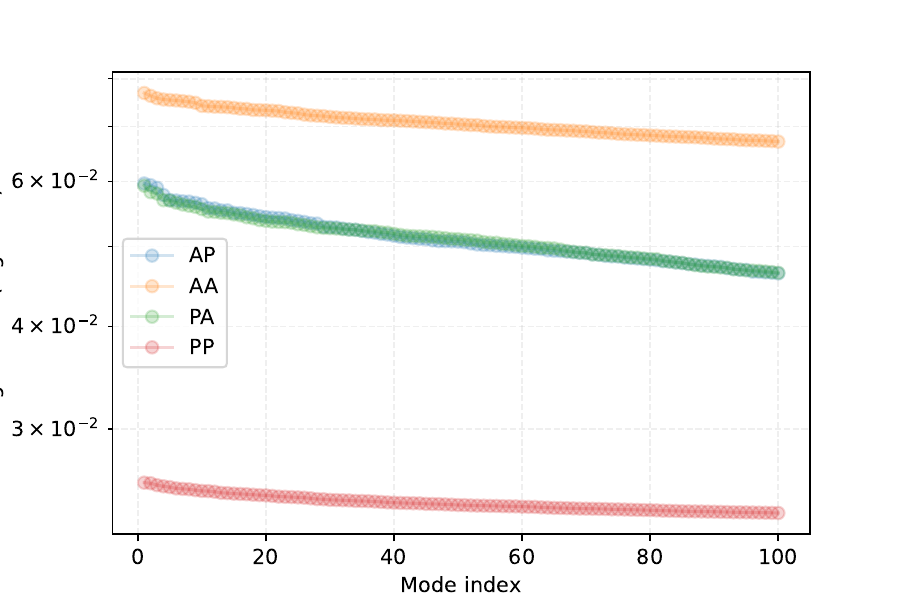}
\caption{The decay of eigenvalues \eqref{eq:effective_coupling} visualized for each sector.}
\label{fig:eigenval}
\end{figure}

\subsection{Modular S-matrix from RBM wavefunction} \label{sec:RBMWave}
We trained the above RBMs on Ising model data with and without spin-flip defect insertions along either the x-cycle, or y-cycle or both. The resulting parameters may be employed to extract ans{\"a}tsze form of the wave functions 
$\Psi^f_a(\tau)=\{\Psi_{aa}(\tau), \Psi_{ap}(\tau),\Psi_{pa}(\tau)\}$ as per \ref{eq:ansatz2}. Their S-transformation gives rise to wavefunctions on $\hat{\Lambda}$ as 
$$\Psi^f_b(-1/\tau) = \{\Psi_{aa}(-1/\tau), \Psi_{ap}(-1/\tau),\Psi_{pa}(-1/\tau)\}\;.$$ 
The inner product between them is given by 
\begin{equation}
O_{ba}=\langle \Psi_b^f(-1/\tau)  | \Psi_a^f(\tau)\rangle\,.
\end{equation}
We have checked numerically small lattices such as \(3\times 3\) and \(4\times 4\) that these overlap matrices are related to $\tilde{S}$ in \eqref{eq:tildeS} through linear transformations among $\Psi^f_a(\tau)$. For example, explicitly enumerating microstates for a \(4\times 4\) lattice shows that the bases  $\Psi^x_a=\{ |1_x\rangle,|\psi_x\rangle,
|\sigma_x\rangle\}$ and $\Psi^y_b=\{ |1_y\rangle,|\psi_y\rangle,|\sigma_y\rangle\}$ 
from equation \eqref{eq:xbasisF} and \eqref{eq:ybasisF} are obtained from our wave functions by the linear transformation
\begin{equation}
P=\left(
\begin{array}{ccc}
 -0.39-0.51 i & -0.43-0.077 i & -0.63 \\
 -0.61-0.21 i & 0.066-0.51 i & 0.56 \\
 -0.18+0.38 i & 0.58-0.45 i & -0.54 \\
\end{array}
\right)\;. 
\end{equation}
We may also formally carry out the above enumeration using the RBM partition function \eqref{eq:prbm_visible} for the same lattice. Again, the lattice is small enough that the enumeration over states (the visible nodes \(\mathbf{v}\)) can be explicitly carried out. On doing so and extracting wave functions as above, we find that the wave functions obtained are related to those of \eqref{eq:xbasisF} and \eqref{eq:ybasisF} by the linear transformation.
\begin{equation}
P=\left(
\begin{array}{ccc} -0.41& -0.72 & -0.63 \\
 -0.61-0.51 i & 0.04-0.7 i & 0.38+0.46 \\
 -0.48+0.12 i & 0.68-0.04 i & -0.56+0.04 i \\
\end{array}
\right)\;. 
\end{equation}
Of course, to fully reproduce the expectations from CFT/TQFT, we need to scale the method to more realistic sampling sizes. While the RBM architecture shown here is fully connected, and a naive extrapolation of the above estimate becomes unfeasible, we expect this will be mitigated. Interestingly, adopting a more `physics motivated' architecture seems to yield promising results which we hope to report on soon.

\section{Conclusions}

In this work we developed a machine-learning construction of wavefunctions for the Ising topological quantum field theory starting from the critical two-dimensional Ising model on the torus. First, we reviewed the lattice implementation of topological defects. Spin-flip defects are realized by reversing the sign of the Ising coupling along a non-contractible seam, equivalently imposing $\mathbb{Z}_2$-twisted boundary conditions. 
These defect insertions organize the torus partition functions into sectors naturally associated with the primary fields $1,\psi,\sigma$ of the Ising CFT and with the ground-state sectors of the Ising TQFT.

Second, we used Monte Carlo configurations from these sectors to train Restricted Boltzmann Machines. Although an RBM is a purely statistical ansatz, its learned parameters show clear physical structure. In particular, the effective covariance matrix $WW^T$ reproduces the dominant nearest-neighbor coupling pattern of the Ising model and detects the anti-ferromagnetic sign reversals introduced by spin-flip defect seams. This demonstrates that the RBM captures not only local correlations but also global information associated with non-contractible defect insertions. The analysis of translationally averaged couplings and distance-dependent decay further indicates that the trained RBM retains signatures of translation invariance and critical scale behavior.

Finally, we constructed RBM wavefunctions from the learned probability weights. The square-root prescription realizes the relation 
\( Z=\langle \Psi|\Psi\rangle \)
at the level of spin configurations and provides an explicit bridge between classical critical partition functions and quantum boundary states. Appropriate linear combinations of the RBM states associated with different spin structures yield candidate representatives of the torus ground states $|1\rangle$, $|\psi\rangle$, and $|\sigma\rangle$. Their overlaps give access to the modular $S$-matrix, providing a nontrivial test that the RBM construction reproduces the expected Ising TQFT modular data.

Several directions remain open. First, the present construction focuses primarily on the real positive amplitudes of the wavefunction. A full representation of a microscopic chiral topological phase would also require learning phase information, which is absent from the classical Boltzmann distribution. One possible route is to use the RBM states obtained here as amplitude seeds and then train complex-valued neural-network wavefunctions against a concrete parent Hamiltonian. Second, the treatment of Kramers-Wannier duality defects can be further developed numerically, since non-invertible defects impose more intricate constraints than ordinary symmetry twists. Third, it would be valuable to extend this approach to other rational CFTs and topological orders, such as Potts models, parafermionic theories, and more general string-net phases, where the relation between defect partition functions and bulk modular data is richer.

\section*{Acknowledgments}
SL's work and computational infrastructure was supported by Beijing Natural Science Foundation grant no IS23010. KBF thanks BIMSA for their hospitality, where this work was initiated.
\appendix
\section*{Appendix}
\section{Training RBMs}
\label{app:RBMtrain}
\begin{figure}
\centering
\includegraphics[width=0.5\linewidth]{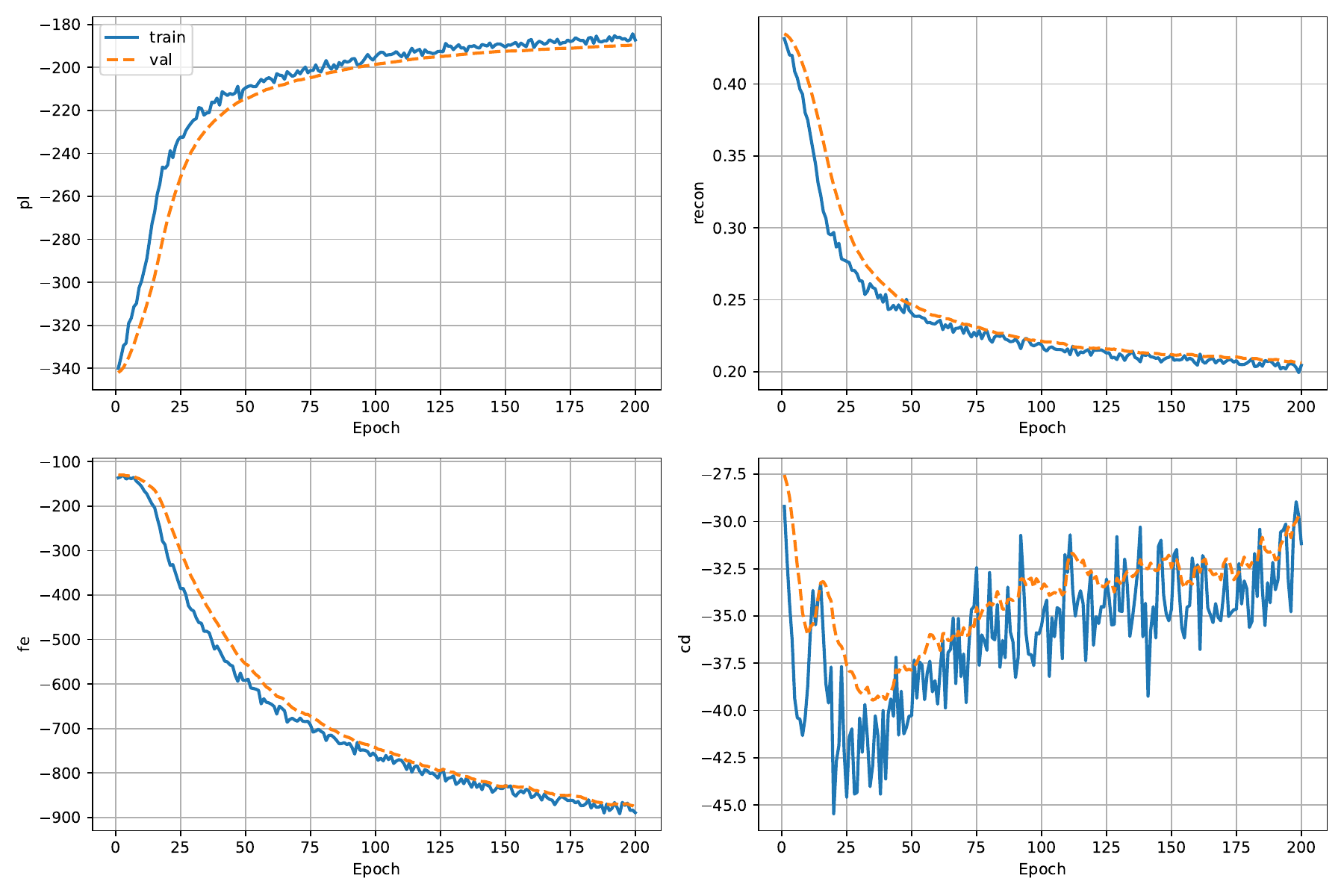}
\caption{Evolution of metrics for training the RBM on the Ising model without defect insertions.}
\label{fig:rbmpptraining}
\end{figure}

\begin{figure}
\centering
\includegraphics[width=0.5\linewidth]{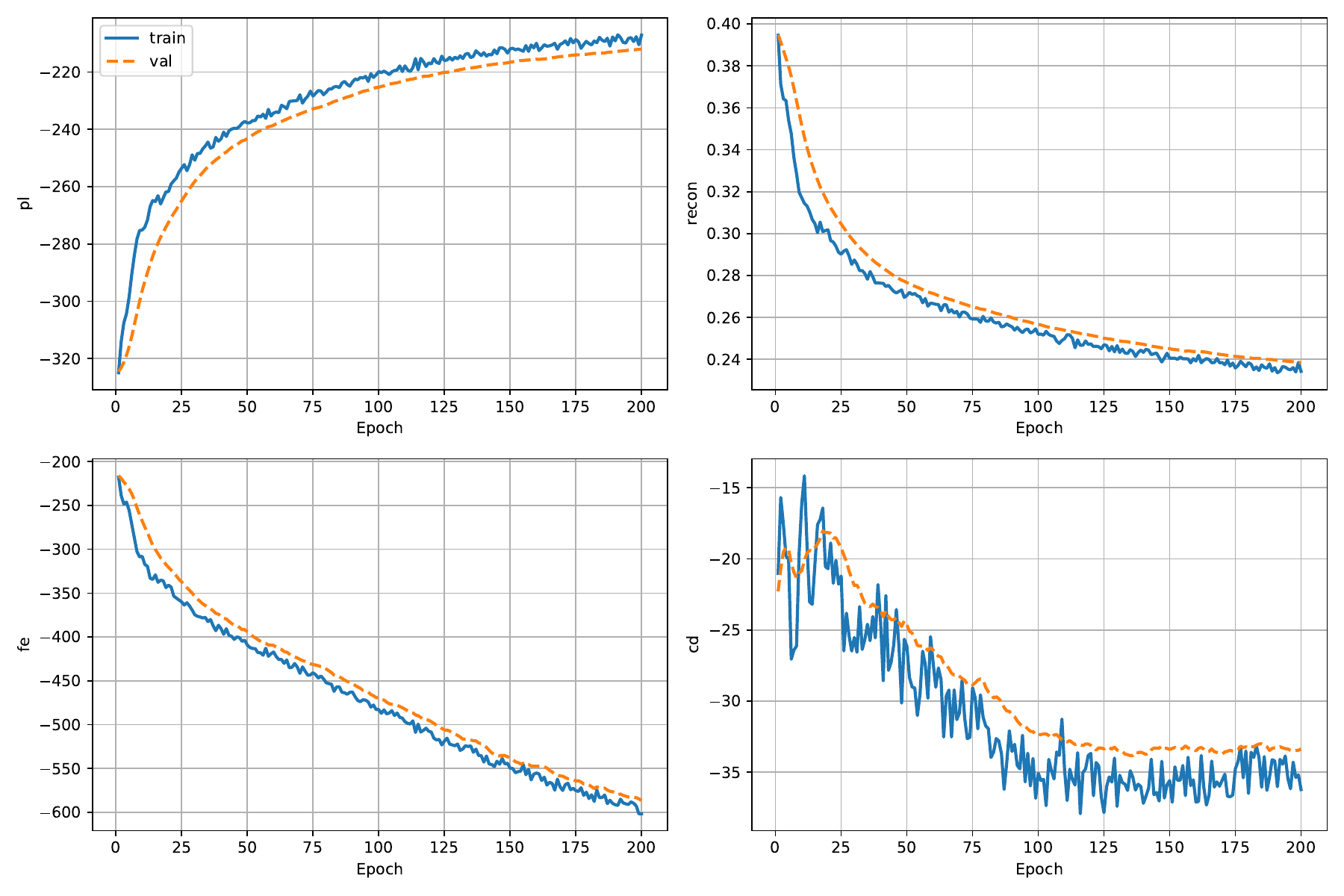}
\caption{Evolution of metrics for training the RBM on the Ising model with 
a spin-flip defect .}
\label{fig:rbmaptraining}
\end{figure}

We train the Restricted Boltzmann Machine (RBM) architecture using a hidden layer dimension of $n_h = 256$. The training ensembles consist of Monte Carlo configurations of the two-dimensional Ising model sampled exactly at the critical coupling $K_c = \frac{1}{2}\ln(1+\sqrt{2})$. To map the topological features of the theory, separate models are trained on snapshots satisfying either untwisted periodic-periodic boundary conditions or configurations with spin-flip defect lines inserted along the non-contractible $a$ and $b$ cycles of the toroidal lattice.
Optimization is performed via stochastic gradient descent (SGD) spanning $200$ epochs, utilizing a constant learning rate of $\eta = 5 \times 10^{-3}$ and a mini-batch size of $64$. To prevent overfitting and promote sparse feature representations, we apply $L_2$ regularization to the weight matrix $\mathbf{W}$ with a regularization strength of $\lambda = 10^{-4}$. Incorporating this weight decay penalizes large parametric amplitudes, effectively shifting the optimization landscape from standard Maximum Likelihood Estimation (MLE) to a Maximum A Posteriori (MAP) framework under a Gaussian prior:
\begin{equation}
\label{eq:map_objective}
\mathcal{L}_{\text{MAP}}(\boldsymbol{\theta}) = \mathcal{L}_{\text{MLE}}(\boldsymbol{\theta}) + \frac{\lambda}{2} \|\mathbf{W}\|_2^2\,.
\end{equation}
Convergence and generalization capabilities are assessed dynamically throughout the optimization schedule by evaluating the log pseudo-likelihood (PL), free energy (FE), reconstruction error ($\mathcal{E}_{\text{recon}}$), and the contrastive divergence (CD) loss proxy across independent training and validation datasets. The empirical evolution of these metrics for both untwisted boundary conditions and a representative spin-flip defect inserted along the imaginary-time cycle are displayed in Figures \ref{fig:rbmpptraining} and \ref{fig:rbmaptraining}, respectively. Technical details for each evaluation metric are outlined below; for an extended treatment of RBM training mechanics, see \cite{hinton2012,fischerRBMintro}.
\subsection{Gradient Evaluation and Sampling Schemes}
\label{app:cd}
Evaluating the gradients $\partial_{\boldsymbol{\theta}}\left[-\log \mathrm{p}_{\boldsymbol{\theta}}(\mathbf{v})\right]$ appearing in \eqref{eq:rbmgradients} requires the simultaneous computation of two expectation values, traditionally designated as the positive and negative phase gradients:
\begin{equation}
\partial^+_{\boldsymbol{\theta}} \equiv \mathbb{E}_{\mathbf{h}\vert \mathbf{v}}\left[\partial_{\boldsymbol{\theta}} E(\mathbf{v},\mathbf{h};\boldsymbol{\theta}) \right]\,, \quad 
\partial^-_{\boldsymbol{\theta}} \equiv \mathbb{E}_{\mathbf{v},\mathbf{h}}\left[\partial_{\boldsymbol{\theta}} E(\mathbf{v},\mathbf{h};\boldsymbol{\theta}) \right]\,.
\end{equation}
The positive phase gradient $\partial^+_{\boldsymbol{\theta}}$ captures data-driven statistics, conditioning the latent representations on the observed empirical configurations. Because the bipartite graph structure forbids intra-layer connections, the conditional distribution factors exactly. Standard probability calculus yields the expected hidden activation conditioned on an empirical data vector $\mathbf{v}$:
\begin{equation}
\mathbb{E}_{\mathbf{h}\vert \mathbf{v}}\left[h_i\right] = \sum_{\mathbf{h}} h_i\,\mathrm{p}\left(\mathbf{h}\vert \mathbf{v}\right) = \sigma(\mathbf{z}_i)\,,
\end{equation}
where $\sigma(\mathbf{z}) = (1 + e^{-\mathbf{z}})^{-1}$ is the element-wise logistic sigmoid function, and the local activation field is given by $\mathbf{z} = \mathbf{v}^T \mathbf{W} + \mathbf{b}^T$.
Conversely, the negative phase gradient $\partial^-_{\boldsymbol{\theta}}$ captures model-driven statistics. Evaluating this expectation exactly requires computing a sum over a $2^{n_v + n_h}$-dimensional configuration space at each optimization step. Consequently, the negative phase must be approximated via Markov Chain Monte Carlo (MCMC) sampling. 
The standard approach relies on alternating Gibbs sampling, alternatingly updating each layer conditioned on the state of the other:
\begin{equation}
\mathbf{h} \sim \mathrm{p}\left(\mathbf{h} \mid \mathbf{v}\right), \qquad
\mathbf{v} \sim \mathrm{p}\left(\mathbf{v} \mid \mathbf{h}\right)\,.
\end{equation}
In Contrastive Divergence ($k$-CD) \cite{hinton2012,hintoncd}, short Gibbs chains are initialized directly at a training data point $\mathbf{v}^{(0)}$ and unrolled for $k$ steps:
\begin{equation}
\mathbf{v}^{(0)} \to \mathbf{h}^{(0)} \to \mathbf{v}^{(1)} \to \mathbf{h}^{(1)} \to \dots \to \mathbf{v}^{(k)}\,.
\end{equation}
This framework can be conceptualized as a product-of-experts architecture \cite{hintoncd}, where individual latent variables act as decoupled feature detectors whose conditional distributions collectively vote on and refine the reconstruction of the visible manifold. The resulting estimator for the negative phase gradients is evaluated at the $k$-th step:
\begin{equation}
\partial^-_{\mathbf{W}_{ij}}E = \mathbf{v}_i^{(k)}\,\mathbf{h}_j^{(k)}\,,\quad 
\partial^-_{\mathbf{b}_j}E = \mathbf{h}_j^{(k)}\,,\quad 
\partial^-_{\mathbf{c}_i}E = \mathbf{v}_i^{(k)}\,.
\end{equation}
Although $k=1$ is often sufficient for empirical convergence \cite{hinton2012,hintoncd}, short chains introduce a systematic bias due to incomplete equilibration.
To suppress this bias, we implement Persistent Contrastive Divergence (PCD) \cite{tieleman2008training}. Rather than re-initializing the Markov chains from empirical data vectors at each gradient step, PCD maintains a persistent set of particles across the entire optimization schedule. Under sufficiently small learning rates, the model parameters shift slowly relative to the mixing time of the chain, ensuring the persistent particles track the true equilibrium distribution of the model and yield less biased, more stable negative-phase expectations.
\subsection{Evaluation Metrics}
Directly monitoring RBM optimization via the exact log-likelihood \eqref{eq:rbm_ansatz_ptheta_v} is computationally intractable due to the presence of the global partition function $\mathcal{Z}(\boldsymbol{\theta})$. To validate convergence and check for overfitting during training, three distinct tractable proxy metrics are evaluated.
\paragraph{Free Energy and Pseudo-Likelihood}
The visible-layer dependence of \eqref{eq:rbm_ansatz_ptheta_v} enters solely through the un-normalized marginal distribution, which can be defined in terms of the free energy $F(\mathbf{v})$:
\begin{equation}
e^{-F(\mathbf{v})} \equiv \sum_{\mathbf{h}} e^{-E(\mathbf{v},\mathbf{h};\boldsymbol{\theta})} = e^{\mathbf{c}^T \mathbf{v}} \prod_{j=1}^{n_h} \left(1 + e^{(\mathbf{v}^T \mathbf{W} + \mathbf{b}^T)_j}\right)\,.
\end{equation}
Next, while $F(\mathbf{v})$ isolates the target data configuration's raw energy coordinate independent of $\mathcal{Z}(\boldsymbol{\theta})$, evaluating global model distributions requires a computable proxy for the full joint likelihood. To this end, we monitor the pseudo-likelihood $PL(\mathbf{v})$, which approximates the joint probability of a configuration $\mathbf{v}$ by assuming conditional independence of its elements:
\begin{equation}
\label{eq:pseudolikelihooddef}
PL(\mathbf{v}) \equiv \prod_{i=1}^{n_v} \mathrm{p}(v_i \mid \mathbf{v}_{\setminus i})\,,
\end{equation}
where $\mathbf{v}_{\setminus i}$ denotes the state of all visible nodes excluding the $i$-th site. To optimize evaluation over large datasets, the product is approximated via stochastic subsampling. Selecting a random index $\hat{k}$, the log pseudo-likelihood is computed as:
\begin{equation}
\label{eq:logPL}
\log PL(\mathbf{v}) \approx n_v \log \sigma\left( F(\mathbf{v}_{\hat{k}\text{-flipped}}) - F(\mathbf{v}) \right)\,,
\end{equation}
where $\mathbf{v}_{\hat{k}\text{-flipped}}$ represents the configuration with its $\hat{k}$-th bit inverted. Successful minimization of the true loss implies that $PL(\mathbf{v}) \to 1$ from below, causing $\log PL(\mathbf{v})$ to asymptote toward zero.
\paragraph{Reconstruction Error:}
A structural assessment of generative fidelity is provided by the Mean Squared Reconstruction Error ($\mathcal{E}_{\text{recon}}$). Given an empirical data vector $\mathbf{v}^{(0)}$, we execute a single-step deterministic Gibbs mapping $\mathbf{v}^{(0)} \to \mathbf{h}^{(1)} \to \mathbf{v}^{(1)}$ via conditional expectations:
\begin{equation}
\mathbf{h}^{(1)}_j = \sigma\left( (\mathbf{v}^{(0)})^T \mathbf{W} + \mathbf{b}^T \right)_j\,, \quad \mathbf{v}^{(1)}_i = \sigma\left( \mathbf{W} \mathbf{h}^{(1)} + \mathbf{c} \right)_i\,.
\end{equation}
The reconstruction error measures the average squared Euclidean distance between the initial data profile and its unrolled reconstruction map:
\begin{equation}
\label{eq:recon_error}
\mathcal{E}_{\text{recon}} = \frac{1}{N n_v} \sum_{n=1}^N \left\| \mathbf{v}_n^{(0)} - \mathbf{v}_n^{(1)} \right\|_2^2\,.
\end{equation}
A decreasing trend in $\mathcal{E}_{\text{recon}}$ signifies that the latent space features adequately capture the relevant correlations necessary to reconstruct the visible data manifold.
\paragraph{Contrastive Divergence Loss Proxy:}
Since optimization updates are governed by the difference between data-driven and model-driven expectations, the actual value of the CD-k gradient can serve as an optimization loss metric. While the full negative phase is intractable, the contrastive divergence scalar proxy ($\mathcal{L}_{\text{CD}}$) evaluates the directional mismatch between the empirical data distribution and the distribution generated after $k$ Gibbs steps:
\begin{equation}
\label{eq:cd_loss_proxy}
\mathcal{L}_{\text{CD}} = \frac{1}{N} \sum_{n=1}^N \left[ F\left(\mathbf{v}_n^{(0)}\right) - F\left(\mathbf{v}_n^{(k)}\right) \right]\,.
\end{equation}
In the early training phase, $\mathcal{L}_{\text{CD}}$ is markedly positive, reflecting that genuine data points occupy significantly lower free energy states than the random states sampled from the initial model distribution. As the network reaches convergence and the generated states $\mathbf{v}^{(k)}$ approach the true empirical distribution, this free energy differential contracts and stabilizes. The evolution of these quantities while training the RBM is shown in Figures \ref{fig:rbmpptraining} and \ref{fig:rbmaptraining} as mentioned previously. We also note the application of annealed importance sampling to estimate 
$\mathcal{Z}(\boldsymbol{\theta })$ for an RBM trained on the Ising model 
\cite{Cossu:2018pxj}.

\bibliographystyle{unsrt}
\bibliography{biblio.bib}
\end{document}